\documentclass{article}
\usepackage[a4paper, left=1.5cm, right=1.5cm, top=2.5cm, bottom=2.5cm]{geometry}
\usepackage{graphicx} % Required for inserting images
\usepackage{amsmath}
\usepackage{amssymb}
\usepackage{authblk}
\usepackage{cite}
\title{Cosmological Implications of Thermodynamic Split Conjecture}
\author{Oem Trivedi \thanks{oem.trivedi@vanderbilt.edu}}
\affil{Department of Physics and Astronomy, Vanderbilt University, Nashville, TN 37235, USA}
\date{\today}

\begin{document}

\maketitle

\begin{abstract}
Building on initial work on the Thermodynamic Split Conjecture (TSC), which posits that black hole and cosmological horizon thermodynamics are generically inequivalent, we examine the consequences of that split for the Gibbons–Hawking temperature and its role across cosmology. We consider many key results in both early and late universe cosmology and show that many important results such as those governing eternal inflation, vacuum tunneling, quantum breaking and primordial black holes can change. The analysis further reveals that small, TSC motivated corrections to horizon thermodynamics can subtly modify Friedmann dynamics, potentially helping to address the $H_0$ and $S_8$ tensions. The work thus provides a unified route from quantum gravity motivated thermodynamics to observational cosmology and motivates dedicated tests of the thermal laws governing the Universe itself.
\end{abstract}

\section{Introduction}
One will not be wrong to say that the current age of cosmology is possibly the most exciting it has ever been, with both theory and observation advancing in an unprecedented synergy to probe the Universe across extraordinary ranges of scale and precision. The discovery and modeling of dark energy \cite{de1SupernovaSearchTeam:1998fmf,de2Li:2012dt,de3Li:2011sd,de4Mortonson:2013zfa,de5Frusciante:2019xia,de6Huterer:2017buf,de7Vagnozzi:2021quy,de8Adil:2023ara,de9Feleppa:2025clx,de10DiValentino:2020evt,de11Nojiri:2010wj,de12Nojiri:2006ri,de13Trivedi:2023zlf,de14Trivedi:2022svr,de15Trivedi:2024inb,de16Trivedi:2024dju} and dark matter \cite{dm11rubin1970rotation,dm1Cirelli:2024ssz,dm2Arbey:2021gdg,dm3Balazs:2024uyj,dm4Eberhardt:2025caq,dm5Bozorgnia:2024pwk,dm6Misiaszek:2023sxe,dm7OHare:2024nmr,dm8Adhikari:2022sbh,dm9Miller:2025yyx,dm10Trivedi:2025vry} have opened vast theoretical landscapes, while persistent anomalies such as the $H_0$ and $S_8$ tensions \cite{ht1DiValentino:2021izs,ht2Clifton:2024mdy,s81kazantzidis2018evolution,s82amon2022non,s83poulin2023sigma,s84Ferreira:2025lrd,baoFerreira:2025lrd} continue to challenge the standard cosmological model and invite radical rethinking of cosmic dynamics. At the same time, a new observational era is dawning, with next–generation CMB and LSS surveys, gravitational–wave observatories, and precision high redshift experiments offering ever tighter constraints on the physics of the early Universe \cite{t1jha2019next,t2Shanks:2015lda,t3chandler2025nsf,t4Euclid:2024yrr,t5WST:2024rai,t6CosmoVerseNetwork:2025alb,t7COMPACT:2022gbl}. This amalgamation of theory and observation has refined not only our quantitative understanding of cosmological parameters but also the very conceptual questions we ask about spacetime and gravity on the largest scales.
\\
\\
Significant attention has hence turned towards the role of quantum gravity in cosmology, seeking to understand how microscopic gravitational degrees of freedom may shape the Universe’s macroscopic evolution. In this context, the recently proposed Thermodynamic Split Conjecture (TSC) \cite{tscpaper} introduces a new way of thinking about cosmic horizons by proposing a fundamental distinction between the thermodynamic structure of black hole event horizons and that of cosmological horizons. The conjecture argues that the Bekenstein–Hawking (BH) entropy \cite{sb1hawking1972black,sb3bekenstein1974generalized,sb4Cotler:2016fpe,sb5krolak1978singularities,sb8joshi2002cosmic,sb9senovilla20151965,sb12hawking1970singularities,sb16janis1968reality} which is rigorously derived in string theory for stationary black holes, cannot be transplanted into an expanding cosmological background without modification since the key equilibrium and boundary conditions that secure its validity are absent in FLRW spacetimes. This conceptual "split" tells us that the thermodynamic properties of cosmology, pertaining to not only its entropy but also temperature and energy flux laws, must be determined intrinsically rather than borrowed from the black hole case. This could be leading towards a new way which we may call as cosmology–native thermodynamics.
\\
\\
While the initial development of the TSC focused on showing that black hole entropy cannot be derived for cosmological horizons from first principles within string theory or quantum gravity, the conjecture’s implications reach much further. In particular, it casts serious doubt on all quantities and arguments in cosmology that depend upon the black hole entropy identification. The most prominent among these is the Gibbons–Hawking (GH) temperature and the related de Sitter thermodynamic analogies. If the TSC holds true, the GH temperature may not represent the actual thermal state of cosmological horizons and consequently, many of the predictions built upon it, ranging from eternal inflation and vacuum stability to bounds like the Trans Planckian Censorship, may require re–evaluation. This recognition motivates the present work and we structure it as follows. In section 2 we give a short overview of the TSC, while in section 3 we discuss what are the issue of the TSC with the Gibbons-Hawking temperature. In section 4 we go into the details of the implications of TSC on early universe cosmology consequently and in section 5 we discuss some related issues beyond the GH temperature and the early universe. We then conclude our work in section 6. 
\\
\\
\section{Thermodynamic Split Conjecture}
The Thermodynamic Split Conjecture \cite{tscpaper} represents a reformulation of how thermodynamic reasoning should be applied to cosmological horizons. It comes from the recognition that the ingredients enabling the success of black hole entropy calculations in string theory are absent in expanding cosmologies. While in black holes, one sees that the Bekenstein–Hawking entropy emerges naturally from controlled microstate counting or gravitational path integral techniques \cite{sv1Strominger:1996sh,sv2Nishioka:2009un,sv3Strominger:1997eq,qe1Sen:2008vm},the same logic cannot be transplanted to cosmology without loss of rigor and continuity. The conjecture therefore says that in any UV complete theory of quantum gravity, the microscopic and thermodynamic structures associated with black hole event horizons and cosmological event horizons are generically inequivalent.
\\
\\
At its core this conjecture arises from a comparison of the boundary conditions, equilibrium structures, and near–horizon universality that make the area law valid in black hole physics. String theory provides several precise derivations of $S_{\rm BH}=A/4G_N$ using D–brane microstate counting \cite{sv4Bena:2005va,sv5Carlip:1998wz,sv6DeHaro:2019gno,sv7Emparan:2006it,car1Cardy:1986ie,car2Cardy:1989ir}, AdS/CFT duality \cite{adscftMaldacena:1997re}, the Quantum Entropy Function \cite{qe2Sen:2009vz,qe3Sen:2012kpz,qe4Banerjee:2010qc,qe5Banerjee:2011jp,qe6Sen:2012cj,qe7Dabholkar:2014ema,qe8Castro:2009jf}, and the Wald entropy formalism. All of these methods rely crucially on features such as asymptotic boundaries supporting conserved charges, timelike Killing horizons ensuring global Gibbs ensembles and universal near–horizon throats that define regulator–independent path integrals. Cosmological spacetimes such as FLRW, various types of Bianchi, dS etc. lack all three as they possess no global boundary to define charges, no global timelike Killing vector to ensure equilibrium and no AdS$_2$–like throat with fixed moduli. Hence, the very framework that produces $S_{\rm BH}=A/4G_N$ for black holes collapses for cosmologies, motivating the idea here that cosmological and black hole thermodynamics represent distinct sectors of gravitational microphysics.
\\
\\
Formally, this inequivalence can be captured through the BKE criterion as well, which introduces a structural triad $(B,K,E)$ representing three essential ingredients. The first element $B$ corresponds to the existence of a boundary or conserved charges defining superselection sectors in the Hilbert space. For a spacetime $(M,g)$ with a future event horizon $H^+$, one writes the Gauss–law fluxes as
\begin{equation}
Q_i = \oint_{S^{d-2}_\infty} *J_i, \qquad \frac{dQ_i}{dt}=0
\end{equation}
and sets $B=1$ if such charges exist and label stable sectors $\mathcal{H}=\oplus_{\{Q_i\}}\mathcal{H}_{\{Q_i\}}$, otherwise $B=0$. The second element $K$ requires a global timelike Killing vector $\chi^a$ that becomes null on $H^+$ and defines a KMS thermal state with inverse temperature $\beta=2\pi/\kappa$, where $\kappa$ is the surface gravity given as
\begin{equation}
\nabla_{(a}\chi_{b)}=0, \qquad \chi^2|_{H^+}=0, \qquad 
\langle O(t+i\beta)O(0)\rangle=\langle O(t)O(0)\rangle
\end{equation}
This ensures that a Gibbs ensemble and Tolman redshift relation are globally well defined and then we finally have the third element $E$ which demands a universal near–horizon region whose geometry and boundary conditions give us a regulator independent macroscopic entropy functional, which is
\begin{equation}
S_{\rm macro}=\frac{A(H^+)}{4G_N}+S_{\rm Wald}^{({\rm higher})}+S_q
\end{equation}
This can be obtained for instance from an AdS$_2\times S^{d-2}$ throat through the Quantum Entropy Function. The BKE criterion then is the statement that if $B\cdot K\cdot E \neq 1$ then there does not exist an observer-independent ensemble $\Omega(E,\{Q_i\}$ such that $S_{\rm micro}=\log\Omega(E,\{Q_i\})=\frac{A(H^+)}{4G_N}+O(A^0)$.
The whole idea hence is that the equality $S_{\rm micro}=S_{\rm macro}$ holds true only when all three conditions are satisfied simultaneously.
\\
\\
Black holes in string theory satisfy $B=K=E=1$ as they possess conserved charges at infinity, stationary Killing horizons and universal near–horizon AdS$_2$ throats where attractor equations fix moduli and the QEF yields $S_{\rm micro}=S_{\rm macro}$. Cosmological spacetimes, however, have $B=K=E=0$ as in FLRW geometries, there is no asymptotic sphere for defining ADM or brane charges, hence $B=0$. There is also no global Killing vector or KMS state, hence $K=0$ and no stationary near–horizon throat or attractor, so $E=0$. Consequently, the black hole style microstate equality cannot even be formulated for cosmology and the same logic can also be applied to global de Sitter space, where the static patch admits an observer–dependent KMS state but not a global one and so therefore, even though a GH temperature $T=H/2\pi$ can be defined locally, it does not correspond to a universal ensemble or conserved charge sector.  
\\
\\
From this criterion, the TSC implies that the entropy associated with cosmological horizons cannot be assigned a black hole like microscopic interpretation and the “area term” seen in semiclassical treatments arises from relational or entanglement constructs across causal patches, not from the degeneracy of charge labeled microstates. Consequently, while black hole and cosmological horizons both exhibit area dependence in semiclassical gravity, the physical origins of their entropies are categorically distinct and could have possibly different scalings too.
\\
\\
There are also two ways to falsify this conjecture, although they are quite difficult to pursue. One route is the bulk microstate geometry path, wherein if one can construct within a single UV complete theory an infinite family of smooth, horizonless cosmological microstate geometries $\{M_\alpha\}$ indistinguishable from a given cosmological background outside a compact region, labeled by invariant indices $I(\alpha)$, and whose degeneracy grows as
\begin{equation}
\text{deg}\,\{M_\alpha\}\sim \exp\!\left[\frac{A(H_{\rm cos})}{4G_N}\right]
\end{equation}
while simultaneously reproducing a unitary Page curve for cosmological radiation with the same leading area coefficient, then a genuine observer–independent ensemble $\Omega(E,\{Q_i\})$ would certainly exist for a cosmological horizon, contradicting the TSC.  
\\
\\
The second route is the holographic path, which is that if one can establish a UV complete cosmological holography with conserved asymptotic charges, a global KMS state, and a regulator–independent macroscopic entropy functional such that
\begin{equation}
B=K=E=1,\qquad 
S_{\rm micro}(E,\{Q_i\})=\log\!\dim\mathcal{H}_{\rm dual}(E,\{Q_i\})
=\frac{A(H_{\rm cos})}{4G_N}+c_0\log A+O(A^0)
\end{equation}
then the conjecture would again be falsified. Either realization would produce a cosmological setting that satisfies the black hole pillars $(B,K,E)=(1,1,1)$ and thus demonstrate the universality of the area law beyond black holes.
\\
\\
Note that in its current form the TSC serves not as an obstruction but as a guidepost as it highlights that cosmological thermodynamics must develop its own structural analogues to $(B,K,E)$. It points towards the possibilities that quasilocal invariants or screen charges could replace their asymptotic counterparts, covariant non-equilibrium statistical frameworks \cite{nonst1sohl2015deep,nonst2gallavotti2019nonequilibrium,nonst3Motta:2025xli,nonst4zwanzig2001nonequilibrium} might be helpful to replace global Gibbs ensembles etc. From this view one sees that the conjecture opens a pathway to a more faithful quantum–gravitational account of cosmic entropy, where one does not assume the black hole analogy but tests it, both theoretically and observationally, as a falsifiable proposition.
\\
\\
\section{TSC's Conflict with Gibbons-Hawking Temperature}
The starting point for much of the literature on cosmological thermodynamics is the observation that de Sitter space possesses a cosmological horizon which is in many ways analogous to the event horizon of a black hole. By importing the BH entropy formula, one associates to the dS horizon an entropy of the form
\begin{equation}
S_{\rm dS} \sim \frac{A}{4G} \sim \frac{\pi}{G H^2}
\end{equation}
where $A = 4\pi H^{-2}$ is the area of the dS horizon and $H$ is the Hubble constant. This relation follows directly by substituting the horizon radius $r_H = H^{-1}$ into the BH area law. Once such an entropy is assumed, one is led to a temperature form for the cosmological horizon
\begin{equation}
T_{\rm GH} \sim \frac{H}{2\pi}
\end{equation}
which is the much celebrated Gibbons-Hawking temperature \cite{gh1Gibbons:1977mu} and so the thermal properties of de Sitter space are, in this construction, directly inherited from the BH entropy of black holes, transplanted to a cosmological setting.
\\
\\
The TSC, however, says that this very transplantation is not justified  and therefore, the very foundation that secures the black hole entropy formula does not exist in cosmology and yet the GH derivation assumes it as a starting axiom. This leads to a sharp tension as the GH temperature is derived by assuming that the cosmological horizon obeys the BH entropy. If the entropy functional in cosmology is instead given by some more general $S_{\rm cos}(H)$, then the link between Euclidean periodicity and horizon thermality does not reproduce the naive $T_{\rm GH}=H/2\pi$. What is meant by that is the GH temperature is not a universal truth but an artifact of importing a black hole law into a setting where its structural justification is absent. The TSC therefore forces one to view the GH temperature as an approximation which is only valid when the cosmological horizons mimic black hole horizons (which may never be completely true), but not a fundamental thermodynamic law of FRW spacetimes.
\\
\\
To illustrate this in detail, let's recall clearly how the derivation of the GH temperature originates from the Euclidean path integral formulation of quantum gravity. In the black hole case, the Euclidean continuation of the Schwarzschild metric 
\begin{equation}
ds^2 = \left(1-\frac{2GM}{r}\right)d\tau^2 + \left(1-\frac{2GM}{r}\right)^{-1}dr^2 + r^2 d\Omega_2^2
\end{equation}
contains a coordinate singularity at the horizon $r=2GM$ and this singularity is removed if the Euclidean time coordinate $\tau$ is made periodic with period
\begin{equation}
\beta = \frac{1}{T_H} \sim 8\pi GM
\end{equation}
this goes to ensure regularity of the Euclidean manifold at the horizon. This compactification leads to a well-defined semiclassical partition function $Z(\beta)=\mathrm{Tr}\,e^{-\beta H}$ and reproduces the Hawking temperature \(T_H = (8\pi GM)^{-1}\). By analogy, the same reasoning is applied to the de Sitter metric
\begin{equation}
ds^2 = -\left(1-H^2r^2\right)dt^2 + \left(1-H^2r^2\right)^{-1}dr^2 + r^2 d\Omega_2^2
\end{equation}
whose Euclidean continuation becomes a 4-sphere of radius \(H^{-1}\) and requiring regularity at the coordinate origin again demands periodic identification of Euclidean time, giving us \cite{gh2Gibbons:1976ue,gh3Gibbons:1978ac,gh4Hawking:1995fd}
\begin{equation}
\beta_{\rm GH} = \frac{2\pi}{H}, \qquad T_{\rm GH} \sim \frac{H}{2\pi}
\end{equation}
This temperature can also be obtained by computing the response of an Unruh-DeWitt detector following a static trajectory in de Sitter space, which registers a thermal spectrum at \(T_{\rm GH}\). The two derivations both give the same thing because both assume that dS space possesses a thermal partition function of the form
\begin{equation}
Z(\beta) = e^{-I_E} \;\; \propto \;\; \exp\!\big(S_{\rm dS} - \beta E\big)
\end{equation}
where \(I_E\) is the Euclidean action and \(S_{\rm dS}\) is identified with the BH entropy \(S_{\rm dS}=A/4G=8\pi^2 M_{\rm Pl}^2/H^2\). This tells us that the statistical interpretation of \(Z(\beta)\) as a canonical partition function and the very normalization of \(T_{\rm GH}\) rely critically on the assumption that dS entropy obeys the same area law as a black hole horizon.
\\
\\
The TSC challenges this step at its foundation as it says that the area-entropy identification is not a theorem in cosmology because the key structural elements that make the black hole case valid, that being asymptotic Killing horizons, conserved charges, and an equilibrium throat separating two static regions, are all absent in an expanding cosmology. If $S_{\rm cos}(H)\neq A/4G$, the Euclidean functional integral does not possess a canonical periodicity in imaginary time that corresponds to a true thermodynamic temperature and so the factor $\beta_{\rm GH}=2\pi/H$ then lacks a physical justification as it is a coordinate artifact of the analytic continuation, not a genuine inverse temperature. From the TSC viewpoint the Euclidean 4-sphere remains a valid geometric continuation sure, but the identification of its period with a physical temperature is not properly justified unless the cosmology-native entropy law $S_{\rm cos}(H)$ empirically reproduces the area law.
\\
\\
One may also note that the Friedmann equations can be re-derived by invoking the Clausius relation at the horizon, $\delta Q = T dS$, with $T$ identified as the GH temperature and $S$ taken to be the BH entropy of the apparent horizon. This line of reasoning has been extensively developed in the literature, but it is only consistent if the cosmological horizon entropy indeed equals $A/4G$ and if the associated temperature is $H/2\pi$. Under the TSC, this closure is accidental and it reflects the internal consistency of assuming GR-derived (Misner-Sharp energy etc.) constructs alongside Black Hole centric constructs in thermodynamics and would not really be an independent proof of the horizon temperature. If the cosmological entropy is in fact different from $A/4G$, as the TSC proposes then the Friedmann equations are recovered only with additional correction terms, or with a non-equilibrium entropy production term. This is something we shall discuss in detail in section 5 and this goes to reinforce the point that the GH temperature is tightly bound to a specific entropy assumption that may not hold in cosmology.
\\
\\
The conflict between the TSC and the GH picture is really direct and structural as the GH result is derived by assuming that the very form of entropy which TSC prohibits for cosmology. If one is to seriously consider the TSC, then this has quite some strong implications as then the GH temperature cannot be taken as a fundamental cosmological law and its ubiquity in inflationary and dark energy models would not reflect not a deep principle of nature but the extension of a black hole result into a domain where it may not apply. The ramifications of such a realization go far and beyond, and we shall be discussing that now.
\\
\\
\section{Implications for Early Universe Cosmology}
After establishing that the TSC calls into question the validity of the GH temperature, it is essential for one to revisit the central constructions of early universe cosmology where this temperature has played a pivotal role. In each case the standard derivation assumes that the horizon entropy is of the BH form and that the corresponding bath temperature of dS space is given by $T_{\rm GH} = H/2\pi$. Once these assumptions are replaced by the more general framework suggested by the TSC, perhaps with empirical parameters determining the effective temperature, entropy, and noise kernels, the conclusions of the standard picture can be substantially altered. In that regard, we consider \begin{equation*}
S_{\rm cos}(H), \qquad T_{\rm eff}(H), \qquad D_{\rm eff}(H)
\end{equation*}
which represent, respectively, the cosmology–native horizon entropy, the effective horizon temperature, and the effective diffusion coefficient governing the coarse grained stochastic dynamics of long–wavelength modes. These functions are to be determined observationally rather than fixed by analogy to black hole thermodynamics, which is in general what is done. However, one useful way to provide a general parametrization for this could be
as follows
\begin{equation*}
T_{\rm eff}(H) = \chi(H)\frac{H}{2\pi}, \qquad S_{\rm cos}(H) = \alpha H^{-p}, \qquad D_{\rm eff}(H) = \beta(H)\frac{H^3}{8\pi^2}
\end{equation*}
where $(\chi, \alpha, p, \beta)$ encode possible departures from the naive BH identifications. Let's now understand the reason these generalized expressions arise too and it all boils down to the links. In the near dS limit, the only dimensionful quantity characterizing the horizon is the Hubble parameter $H$, so any effective temperature must scale  with $H$ up to a dimensionless prefactor. This motivates the general form as follows
\begin{equation} \label{te}
T_{\rm eff}(H) = \chi(H)\frac{H}{2\pi}
\end{equation}
where $\chi(H)$ encodes possible departures from exact GH thermality as we have discussed so far. If $\chi=1$, the standard result is recovered while $\chi(H)\neq 1$ could reflect possible quantum gravity corrections to horizon thermodynamics.
\\
\\
The scaling of entropy is more subtle as in the black hole case, the entropy is an area law $S\propto A\propto H^{-2}$ but without the structural features that guarantee this in black holes, the scaling exponent need not really be two as discussed in \cite{tscpaper}. Since the horizon radius scales as $H^{-1}$, a simple yet general power law form may be
\begin{equation} \label{scos}
S_{\rm cos}(H) = \alpha H^{-p}
\end{equation}
with $\alpha$ a dimensionful coefficient and $p$ an exponent to be fixed by cosmological observation. The BH choice corresponds to $(\alpha,p)=(8\pi^2 M_{\rm Pl}^2,2)$, but nothing in FRW spacetime enforces this.
\\
\\
One is thus led here to the notion of a cosmology–native entropy, which is defined not by direct transplantation of black hole relations but by thermodynamic principles applied intrinsically to FRW spacetimes. The proposal advanced in \cite{tscpaper} is precisely to treat the horizon entropy as an empirical law, measurable through consistency relations between energy flux, expansion rate, and horizon dynamics. In this approach, the entropy scaling is not fixed by area but is instead inferred from the generalized Clausius relation $\delta Q = T,dS$ applied to the cosmological horizon with $T=T_{\rm eff}(H)$ and $S=S_{\rm cos}(H)$. The key shift is conceptual as the entropy becomes a function to be determined by cosmological observation, with the Bekenstein–Hawking area law recovered only if the data prefer $p=2$ and so in this sense, $S_{\rm cos}(H)$ encodes the true thermodynamic evolution of the universe, grounded in horizon dynamics rather than imported from black hole analogies.
\\
\\
Establishing the form of $S_{\rm cos}(H)$ observationally requires data that probe horizon–scale fluctuations across a broad range of redshifts and this is exactly where 21cm intensity mapping becomes indispensable, like from those assembled from SKA \cite{ska1Weltman:2018zrl,ska2Maartens:2015mra,ska3Santos:2015gra}. Unlike the CMB which fixes a single early epoch and galaxy surveys which access relatively late times, the 21cm signal traces neutral hydrogen across much of cosmic history, potentially from $z\sim 30$ down to reionization \cite{hera1DeBoer:2016tnn,hera2HERA:2021noe} and beyond. The statistical properties of these fluctuations provide a direct window into horizon scale thermodynamics and by measuring the variance and noise kernels of super–Hubble modes, one can then infer the effective diffusion constant $D_{\rm eff}(H)$. From this one can then reconstruct both $T_{\rm eff}(H)$ and $S_{\rm cos}(H)$ and in this way, 21cm mapping acts as a thermodynamic interferometer for cosmology which would enable the first empirical calibration of the entropy law that governs horizons in an expanding universe.
\\
\\
Finally, the stochastic noise driving super-Hubble fluctuations in inflation is commonly tied to the GH temperature, yielding the variance $(H/2\pi)^2$ per Hubble time and a diffusion constant $D=H^3/8\pi^2$ but if the bath temperature and entropy law differ, then the corresponding fluctuation-dissipation relation acquires a correction factor. We can say that this factor could be some $\beta(H)$, which accounts for deviations from white-noise statistics and modified KMS periodicity and this can then lead us to
\begin{equation} \label{de}
D_{\rm eff}(H) = \beta(H)\frac{H^3}{8\pi^2}
\end{equation}
Note that the standard picture corresponds to $\beta=1$, while $\beta(H)\neq 1$ captures colored or suppressed noise characteristic of a non-GH bath.
\\
\\
With these definitions cleared, we now head over to understanding the TSC implications on cosmology and in this line, consider first the case of eternal inflation. The stochastic inflation formalism describes a long-wavelength inflaton field as a coarse-grained variable evolving under a Langevin equation and splitting the field into short and long modes with respect to the Hubble scale, the long-wavelength inflaton obeys
\begin{equation}
\dot{\phi}_{\ell}(t) = -\frac{V'(\phi_\ell)}{3H} + \xi(t)
\end{equation}
where the first term is the classical slow-roll drift and $\xi(t)$ is a stochastic noise source generated by the continual horizon-crossing of short modes. The correlation function of the noise is given as
\begin{equation}
\langle \xi(t)\xi(t') \rangle = \frac{H^3}{4\pi^2}\,\delta(t-t')
\end{equation}
so that the inflaton undergoes a Brownian motion with variance
\begin{equation}
\delta\phi_q^2 \sim \left(\frac{H}{2\pi}\right)^2
\end{equation}
per Hubble time and this identification is rooted in the assumption that the horizon crossing modes thermalize with a bath of effective temperature $T_{\rm GH}=H/2\pi$, which arises from the periodicity of Euclidean de Sitter space. The quantum diffusion term in the Langevin equation is thus directly tied to the GH temperature.
\\
\\
At the same time, the inflaton experiences a classical drift per Hubble time given by
\begin{equation}
\delta\phi_c \sim \frac{|V'|}{3H^2}
\end{equation}
Eternal inflation thus occurs when the random quantum step exceeds the deterministic classical drift, that is \cite{et2gibbons1983very,et3vilenkin1983birth,et4guth1985quantum,et5vanchurin2000predictability,et6linde1986eternal,et7linde1986eternally,et8goncharov1987global,et9trivedi2022rejuvenating,et10guth2000inflation,et1guth2007eternal}
\begin{equation}
\frac{H^2}{2\pi} \gtrsim \frac{|V'|}{3H^2}
\end{equation}
The inequality here implies that sufficiently flat potentials always lead to self reproducing patches of inflation in the standard picture and this creates the possibility of a "Multiverse". Under the TSC, however, the use of the GH temperature to normalize the noise kernel is not justified and so the effective diffusion coefficient must instead be inferred empirically, and can be parameterized as
\begin{equation}
D_{\rm eff}(H) = \beta(H)\frac{H^3}{8\pi^2}
\end{equation}
The variance per Hubble time is then modified relative to the GH result, and the eternal inflation condition becomes
\begin{equation}
\beta(H)\frac{H^4}{(2\pi)^2} \gtrsim \frac{V'^2}{9H^4}
\end{equation}
This is quite an interesting point, as when $\beta(H)<1$ the noise is suppressed compared to the GH expectation and many potentials that previously supported eternal inflation cease to do so. Conversely if $\beta(H)>1$, diffusion is enhanced and eternalization becomes more generic. So, the crucial point is that the very existence of eternal inflation is not dictated by inflationary dynamics alone, but hinges on the assumption of GH thermality and with the TSC framework, eternal inflation is no longer a robust prediction but an empirical question tied to the true thermodynamics of cosmological horizons. Let's also now do this analysis in a general way, not assuming the forms \eqref{te}-\eqref{de}. The dynamics of long wavelength inflaton modes can again be described by a Langevin equation
\begin{equation*}
\dot{\phi}_\ell = -\frac{V'(\phi_\ell)}{3H} + \xi(t), \qquad \langle \xi(t)\xi(t') \rangle = N(t,t')
\end{equation*}
The cumulative variance per Hubble time is
\begin{equation}
D_H(H) = \int_t^{t+H^{-1}} dt_1 \int_t^{t+H^{-1}} dt_2\, N(t_1,t_2)
\end{equation}
this reduces to $D_H \simeq 2D_{\rm eff}(H)/H$ for approximately local noise and the quantum jump per Hubble time is $\delta\phi_q^2\simeq D_H(H)$, while the classical drift is $\delta\phi_c\simeq |V'|/(3H^2)$. The onset of eternal inflation follows the condition
\begin{equation}
D_H(H) \gtrsim \frac{V'^2}{9H^4}
\end{equation}
This condition here is now fully general and makes no reference to the GH temperature or the form \eqref{te}. Once $D_{\rm eff}(H)$ is measured from data, the existence or absence of an eternally self reproducing phase is again seen as  an empirical question rather than one fixed by an assumed $T_{\rm GH}=H/2\pi$.
\\
\\
A closely related process is the Hawking--Moss transition, which describes tunneling of the inflaton field across a potential barrier separating two vacua \cite{hm1hawking1982supercooled}. In the Euclidean picture, one evaluates the path integral for field configurations in which the inflaton fluctuates from the local minimum of the potential to the top of the barrier. The dominant saddle in this case is a Euclidean dS solution with the field fixed at the top and the tunneling rate is obtained as the exponential of minus the difference between the Euclidean actions of the two solutions \cite{hm2coleman1980gravitational,hm3gregory2020black,hm4coleman1977fate,hm5callan1977fate}. This difference is proportional to the change in dS entropy but crucially, under the assumption that the horizon entropy is given by the BH formula. In the standard picture one therefore obtains
\begin{equation}
\Gamma_{\rm HM} \propto \exp\!\left[\Delta S_{\rm dS}\right], \qquad 
\Delta S_{\rm dS} = 8\pi^2 M_{\rm Pl}^2\left(\frac{1}{H^2(\phi_{\rm top})}-\frac{1}{H^2(\phi_{\rm min})}\right)
\end{equation}
The appearance of $H/2\pi$ in the underlying Euclidean construction reflects the same GH thermality that is at the heart of stochastic inflation as well and the tunneling rate is fixed by the entropy difference between the horizon in the false vacuum and the horizon at the top of the barrier. This reasoning ties the very existence of vacuum transitions to the use of BH entropy in cosmology.
\\
\\
Given the TSC, the exponent governing the Hawking-Moss transition must be rewritten in terms of the cosmological entropy functional and with
\begin{equation*}
S_{\rm cos}(H) = \alpha H^{-p}
\end{equation*}
the transition rate becomes
\begin{equation}
\Gamma_{\rm HM} \propto \exp\!\left[\alpha\left(H^{-p}(\phi_{\rm top}) - H^{-p}(\phi_{\rm min})\right)\right]
\end{equation}
The stability of vacua is then highly sensitive to the value of the entropy exponent $p$ and this has important ramifications. For $p<2$ the entropy difference between the top and the minimum is reduced compared to the standard case, suppressing tunneling and rendering vacua more stable but for $p>2$ the difference is amplified which leads to enhancing tunneling and making vacua less stable. The long-term dynamics of the inflationary landscape and the notions of whether metastable vacua can survive on cosmological timescales, thus depend not only on the potential shape but also on the empirical entropy law that governs cosmological horizons. Under the TSC this law is not assumed but must be measured, so the fate of the landscape is elevated from a theoretical postulate to an observationally constrained question. Note that in the general case where we do not assume \eqref{scos}, the difference of actions can be expressed as
\begin{equation}
\Gamma_{\rm HM} \propto \exp\!\big[\Delta S_{\rm cos}\big], \qquad 
\Delta S_{\rm cos} = S_{\rm cos}\!\big(H(\phi_{\rm top})\big) - S_{\rm cos}\!\big(H(\phi_{\rm min})\big).
\end{equation}
The stability or metastability of inflationary vacua therefore depends only on the functional form of $S_{\rm cos}(H)$, whatever it maybe, which is determined observationally and this avoids assuming an area law altogether.
\\
\\
Another central result concerns the species and information bounds as conventionally the maximum number of light fields consistent with dS is bounded by the horizon entropy. Considering the heuristic argument that each species consumes at least one bit of entropy, we get
\begin{equation}
N_s \lesssim S_{\rm dS} \sim \frac{M_{\rm Pl}^2}{H^2}
\end{equation}
Under TSC the correct replacement is
\begin{equation}
N_s \lesssim S_{\rm cos}(H) = \alpha H^{-p}
\end{equation}
The tightness of the bound therefore depends directly on $p$ and so if $p<2$, the bound becomes stricter and disfavors theories with many light fields, such as axiverse scenarios and if $p>2$ the bound relaxes, hence opening room for a richer light sector. In the general case the bound becomes \begin{equation}
N_s \lesssim S_{\rm cos}(H)
\end{equation}
where the right hand side would now depend on the empirically determined entropy, not on a pre–chosen power of $H^{-1}$. The conclusion would be that the viability of multi-field inflationary models is thus directly tied to the cosmological entropy law as well. 
\\
\\
The question of quantum breaking time \cite{br1dvali2017quantum} provides yet another example that we need to think about. In the standard case the semiclassical dS state has a lifetime set by
\begin{equation}
t_Q \sim \frac{S_{\rm dS}}{H} \sim \frac{M_{\rm Pl}^2}{H^3}
\end{equation}
which is derived by dividing the de Sitter entropy by the number of quanta produced per Hubble time, assuming GH temperature and BH entropy and so under TSC the estimate becomes
\begin{equation}
t_Q \sim \frac{S_{\rm cos}(H)}{H} = \frac{\alpha}{H^{p+1}}
\end{equation}
So in the general case we see that if the true cosmological entropy differs from the area law, the quantum breaking time will correspondingly lengthen or shorten, again in a measurable way. If we are to consider \eqref{scos} though, then this scaling reveals that if $p<2$, quantum breaking is delayed and inflationary or dark energy phases may last much longer than previously expected while if $p>2$, the lifetime is shorter and cosmic acceleration becomes unstable on relatively fast timescales. The semiclassical consistency of de Sitter space is thus no longer universal in general but contingent on the empirically determined entropy scaling.
\\
\\
The Trans Planckian Censorship Conjecture \cite{tcc1Bedroya:2019snp,tcc2Bedroya:2019tba,tcc3Andriot:2020lea,tcc4Brahma:2020zpk,tcc5Blumenhagen:2019vgj,tcc6Mizuno:2019bxy,tcc7Brahma:2019unn,tcc8Berera:2019zdd,tcc9Cai:2019hge} (TCC) also finds itself reshaped by TSC. The TCC posits that modes that begin below the Planck length cannot be stretched beyond the Hubble radius, leading to the standard bound
\begin{equation}
N_{\rm max} \lesssim \ln\frac{M_{\rm Pl}}{H_{\rm end}}
\end{equation}
which is derived under the assumption that sub-Planckian modes thermalize according to GH temperature and become physical fluctuations. If one applies TSC in this case then the classical nature of the modes does not depend on $H/2\pi$ but on the effective noise and temperature. With this in mind, the revised bound takes the form
\begin{equation}
N_{\rm max}^{\rm TSC} \lesssim \ln\frac{M_{\rm Pl}}{H_{\rm end}} - \ln\mathcal{C}_{\rm cl}(\chi,\beta,\tau_c)
\end{equation}
where $\mathcal{C}_{\rm cl}$ quantifies the difficulty of decoherence with colored noise and suppressed temperature. The general form of the bound can also be written as
\begin{equation}
N_{\rm max} \lesssim \ln\!\frac{M_{\rm Pl}}{H_{\rm end}} - \ln \mathcal{C}_{\rm cl}\!\big[T_{\rm eff}(H),D_{\rm eff}(H),N(t,t')\big]
\end{equation}
where $\mathcal{C}_{\rm cl}$ is now a more general function to quantify the efficiency of decoherence and so this expression remains valid without specifying any functional form for the entropy or temperature. In practice this tends to tighten the censorship and making it more difficult for modes to classicalize.
\\
\\
Primordial Black Hole (PBH) \cite{pbh10Belotsky:2018wph,pbh2Carr:1974nx,pbh3Carr:1975qj} production provides another avenue where the use of GH thermality enters directly. The idea is that during inflation, in particular during phases such as ultra slow roll, the curvature perturbation power spectrum can be temporarily enhanced on small scales. When these amplified fluctuations reenter the horizon in the radiation era, overdense regions above a threshold can collapse to form black holes. The abundance of PBHs is therefore directly tied to the amplitude of field fluctuations generated during inflation \cite{pbh4Carr:2020gox,pbh5Page:1976df,pbh6Carr:2016drx,pbh7Khlopov:2008qy,pbh8Khlopov:1985fch,pbh9Khlopov:1980mg}. If we are to think of the standard stochastic picture, the inflaton field experiences random kicks with variance per Hubble time that we have repeatedly discussed here to be
\begin{equation*}
\langle \delta \phi^2 \rangle \sim \left(\frac{H}{2\pi}\right)^2
\end{equation*}
and this variance feeds directly into the curvature perturbation spectrum, which in turn controls the probability distribution of density contrasts $\delta$ when scales reenter the horizon. The collapse fraction of regions into PBHs is exponentially sensitive to this variance, so even small changes in the fluctuation amplitude can drastically alter predicted PBH abundances.
\\
\\
With the TSC, the assignment of $H/2\pi$ as a genuine bath temperature is no longer justified and so the effective stochastic noise must instead be described by an empirical diffusion constant, which we can parametrize as \eqref{de} and so the variance of the field per Hubble time is then given by
\begin{equation}
\langle \delta \phi^2 \rangle \sim D_{\rm eff}\,\Delta t
\end{equation}
where $\Delta t \sim H^{-1}$ is the coarse-graining timescale and note that in the general case one can write
\begin{equation}
\langle \delta\phi^2 \rangle_{H^{-1}} = D_H(H)
\end{equation}
The small–scale power spectrum $\mathcal{P}_\zeta(k)\propto D_H(H)/\dot{\phi}^2$ then directly depends on the empirically determined diffusion coefficient and so PBH production is enhanced or suppressed accordingly, independent of any assumed $H$–scaling of $S_{\rm cos}$. With the consideration of \eqref{de} the factor $\beta(H)$ thus captures the departure from the idealized GH noise kernel. If $\beta(H)<1$, we have that the fluctuations are suppressed relative to the standard prediction, and this leads  to a dramatic reduction in PBH production but if $\beta(H)>1$, the fluctuations are amplified, which ends up making PBH formation much more efficient.
\\
\\
The key point is that PBH abundance is not a universal prediction of inflationary dynamics alone, but a sensitive diagnostic of the noise kernel assumed for super-Hubble modes. In the standard approach this kernel is fixed by the GH temperature, while under the TSC its form must be measured so the very existence and cosmological significance of PBHs therefore becomes a question of observational thermodynamics of horizons as well, and not an inevitable consequence of inflationary model building.
\\
\\
In each of these cases the lesson is consistent, with that being that the core equations of early universe cosmology rely somehow directly or indirectly on the assumption that the GH temperature and BH entropy apply to cosmological horizons. The TSC calls this assumption into doubt, replacing the fixed $H/2\pi$ and $H^{-2}$ scalings with general exponents and prefactors to be determined by observation. As a result, what once seemed like natural predictions of early universe physics become hypotheses that need to be tested. Eternal inflation, vacuum stability, species bounds, semiclassical lifetimes, trans-Planckian censorship, and primordial black hole production may all require reformulation.
\\
\\
\section{Issues Beyond GH and Early Universe}

The GH construction finds its most widely used dynamical extension in the temperature assigned to the apparent horizon of FRW spacetimes. The apparent-horizon radius is $r_A=(H^2+k/a^2)^{-1/2}$ and a temperature is introduced by analogy with the de Sitter result, with the sign choice dictated by a quasi-static limit of the Kodama-Hayward surface gravity \cite{ha1Hayward:1997jp,ha2Hayward:2008jq}. In exact de Sitter with $k=0$ and $\dot H=0$, that temperature reduces to the GH temperature but the general case,however, we arrive at a different temperature prescription which is known as the Cai-Kim (CK) temperature \cite{Cai:2005ra}, which also deserves special mention here. It is defined on the apparent horizon of FRW spacetimes as
\begin{equation}
T_{\rm CK} \sim \frac{1}{2\pi r_A}, \qquad r_A = \frac{1}{\sqrt{H^2 + k/a^2}}
\end{equation}
In exact de Sitter with $k=0$ this reduces to $T_{\rm GH}=H/2\pi$, but for time-dependent backgrounds it is usually taken as the relevant horizon temperature. The prescription is motivated by the Kodama-Hayward surface gravity, where dynamical corrections of order $\dot r_A/(Hr_A)$ are typically dropped in order to maintain a positive and quasi-static definition. This is why the Cai-Kim form has become the default assignment in discussions of FRW thermodynamics.
\\
\\
The TSC also places this prescription into question in precisely the same way that it challenges the GH result. The assignment of $T_{\rm CK}=1/2\pi r_A$ supposes apriori that the entropy of the apparent horizon is still given by the BH area law
\begin{equation*}
S \sim \frac{A}{4G} \sim \frac{\pi r_A^2}{G}
\end{equation*}
Without this assumption the logical bridge that equates the Kodama-Hayward surface gravity with a physical bath temperature for long-wavelength modes collapses. If the entropy functional is in fact $S_{\rm cos}(H)=\alpha H^{-p}$, then the Cai-Kim temperature cannot remain a universal law but must instead deform into
\begin{equation}
T_{\rm eff}(t) = \chi(H)\,\frac{1}{2\pi r_A}\left(1 - \frac{\dot r_A}{2Hr_A} + \cdots\right)
\end{equation}
where the dynamical correction that is normally discarded reappears as part of the observationally relevant quantity and of course, for $\chi(H)=1$ we recover the usual CK formula. Thus, even though the CK formula is designed to extend $H/2\pi$ to more general spacetimes, it inherits the same vulnerability exposed by the TSC, namely that the identification of horizon entropy with $A/4G$ is not justified in cosmology.
\\
\\
This has further consequences for the many derivations that rely on CK. For instance, it has been shown that the Friedmann equations can be recovered by combining the Clausius relation with $T_{\rm CK}$ and $S=A/4G$. If either ingredient is altered then the equilibrium Clausius law yields not the Einstein equations but their deformed counterparts, or else requires the introduction of non-equilibrium entropy production. The TSC therefore implies that while the CK form may remain a useful approximation in regimes close to de Sitter, its claim to universality across dynamical FRW cosmologies is perhaps not attainable and in this way the CK generalization is not exempt from scrutiny but rather exemplifies how deeply the TSC forces one to reconsider cosmological thermodynamics. 
\\
\\
There are other slightly non-trivial ways in which TSC affects other areas in cosmology. Euclidean path integral is frequently invoked to motivate thermodynamic identifications for horizons \cite{z1Govindarajan:2002ry}, like in the black hole case one analytically continues to Euclidean signature and removes a conical singularity by fixing the Euclidean time period to $\beta=1/T$ and so obtaining a semiclassical partition function $Z\sim e^{-I_E}$ and a Helmholtz free energy $F=-(1/\beta)\ln Z$. In dS space the Euclidean manifold is a four-sphere with radius $H^{-1}$ and the same logic yields $\beta=2\pi/H$ and $T=H/2\pi$. The statistical interpretation assumes that the entropy is given by the BH area law and that the Euclidean action evaluates the correct saddle within a well-defined measure and so even setting aside the conformal-factor problem of Euclidean gravity, in cosmology there is no asymptotic Hilbert space furnishing a canonical notion of energy or a partition function in the standard thermodynamic sense. Then the usual equalities like $S=-\partial F/\partial T$ and $C=-\beta^2 \partial^2 \ln Z/\partial \beta^2$ become formal identities whose physical meaning collapses without the area-entropy input and if one takes the TSC to be true then  the replacement of $A/4G$ by $S_{\rm cos}(H)=\alpha H^{-p}$ invalidates the presumed equality between the Euclidean period and a physical bath temperature for the long modes, so that
\begin{equation}
Z_{\rm cos}\;\not\sim\;\exp\!\Big[+\alpha H^{-p}\Big], \qquad F_{\rm cos}\;\not=\;-\frac{1}{\beta}\ln Z_{\rm cos}, \qquad \beta=\frac{2\pi}{H}
\end{equation}
Note how this is an issue which is more fundamental than just a calculation subtlety. The path integral saddle can still be a useful semiclassical device but its thermodynamic reading is no longer secured unless and until the cosmological entropy law is established independently.
\\
\\
Another interesting arena where entropy law mediates a new paradigm arises in the quantum extremal surface program for cosmological horizons. The generalized entropy used in the island formula is $S_{\rm gen}=A/4G+S_{\rm bulk}$ and quantum extremality of $S_{\rm gen}$ on candidate boundaries selects the island and determines Page-like behavior for the entropy seen by an observer. In cosmology one can define analogous regions anchored to the cosmological horizon and search for quantum extremal surfaces that render the observer's algebra finite and consistent with unitarity, leading to cosmological islands \cite{is1Hartman:2020khs,is2Azarnia:2021uch,is3Espindola:2022fqb,is4Bousso:2022gth,is5Ben-Dayan:2022nmb}. The TSC alters this picture at its hinge point as if the boundary-entropy functional is not $A/4G$ but rather a cosmology native functional tied to $H$ or more general quasi local invariants then the generalized entropy becomes
\begin{equation}
S_{\rm gen}^{\rm cos} \;=\; S_{\rm bdy}^{\rm cos}[H,\ldots] \;+\; S_{\rm bulk}
\end{equation}
and the quantum extremality condition is correspondingly deformed. The position of extremal surfaces, the value of the generalized entropy and the emergent Page time for cosmological observers are all contingent on the correct boundary functional. In the black hole case the area term is justified by microstate counting and semiclassical consistency checks. In cosmology those justifications do not transfer without modification and the island paradigm retains its structural appeal but its quantitative content depends on the same horizon thermodynamic identification that the TSC questions.
\\
\\
There is a different route by which horizon thermodynamics is often tied back to cosmological dynamics. One writes the Clausius relation $\delta Q=T\,dS$ on the apparent horizon and uses the unified first law of Hayward along with the Misner-Sharp energy and the perfect-fluid stress tensor to relate the heat flux to cosmic expansion. With $T=1/2\pi r_A$ and $S=A/4G$ one recovers the Friedmann equations \cite{therm1Khodam-Mohammadi:2023ndu,therm2Nojiri:2022aof,therm3Akbar:2006kj,therm4Bargueno:2021nuc,therm5Sheykhi:2010zz,therm6Helou:2015yqa,therm7Nojiri:2019skr,therm8Nojiri:2022nmu,therm9Nojiri:2022dkr,therm10Odintsov:2022qnn,therm11Nojiri:2023bom,therm12Nojiri:2024zdu,therm13Nojiri:2023ikl,therm14deHaro:2023lbq,therm15Brevik:2024ozg}. But the TSC touches upon an often ignored subtlety here, which is that the unified first law and the Misner-Sharp energy are themselves GR constructs derived from the Einstein equations, while the assignments of $T$ and $S$ are black hole inspired identifications which are imported into FLRW. The agreement is therefore an equivalence within GR but not an independent derivation of GR dynamics from horizon thermodynamics. If instead the entropy and temperature are given by the cosmology-native proxies, the Clausius balance becomes an empirical relation with two possibilities. In the equilibrium reading one sets $\delta Q=T_{\rm eff}\,dS_{\rm cos}$ and obtains a generalized second Friedmann equation. For spatial flatness with $r_A=H^{-1}$ and $k=0$ the heat flow across the horizon is then given as 
\begin{equation}
\frac{dQ}{dt} \;=\; \frac{4\pi}{H^2}\,(\rho+p)
\end{equation}
In the general case we can write
\begin{equation}
\dot Q = T_{\rm eff}(H)\,\frac{dS_{\rm cos}(H)}{dt}
\end{equation}
and differentiating $S_{\rm cos}(H)$ with respect to $H$ and invoking the continuity equation leads to a generalized dynamical relation for $\dot H$
\begin{equation}
\dot H = -\frac{4\pi}{H^2}\,\frac{\rho+p}{\tfrac{1}{2\pi}T_{\rm eff}(H)\,S_{\rm cos}'(H)}
\end{equation}
This form explicitly tells us that all standard cosmological evolution equations and fluctuation amplitudes depend only on the form of functions $T_{\rm eff}(H)$, $S_{\rm cos}'(H)$, and $D_{\rm eff}(H)$ and so the TSC therefore leads to a new direction of thoughts here too. Consider the entropy change under \eqref{scos} now, which is
\begin{equation}
\frac{dS_{\rm cos}}{dt} \;=\; -\,p\,\alpha\,H^{-(p+1)}\,\dot H
\end{equation}
Equating $\dot Q = T_{\rm eff}\,\dot S_{\rm cos}$ with $T_{\rm eff}=\chi(H) H/2\pi$ gives us
\begin{equation}
\dot H \;=\; -\,\frac{8\pi^2}{\chi\,p\,\alpha}\,(\rho+p)\,H^{\,p-2}
\end{equation}
which reduces to the GR relation $\dot H=-4\pi G(\rho+p)$ when $(\chi,p,\alpha)=(1,2,8\pi^2 M_{\rm Pl}^2)$. Combined with the continuity equation this determines a modified $H(a)$ and upon integration, a Friedmann-like constraint of the form 
\begin{equation} \label{fr1}
H^2 \;=\; \frac{8\pi G}{3}\,\rho \;+\; F\!\left(H;\,\alpha,p,\chi\right)
\end{equation}
is achieved, where $F$ vanishes in the BH and GH limit but otherwise acts as an effective source determined by the cosmological entropy and temperature laws. In the non-equilibrium reading one writes instead
\begin{equation} \label{q1}
\dot Q \;=\; T_{\rm eff}\,\dot S_{\rm cos} \;+\; T_{\rm eff}\,\dot S_i, \qquad \dot S_i \ge 0
\end{equation}
which leads to
\begin{equation} \label{fr2}
\dot H \;=\; -\,\frac{8\pi^2}{\chi\,p\,\alpha}\,(\rho+p)\,H^{\,p-2} \;+\; \frac{H^{\,p+1}}{p\,\alpha}\,\dot S_i
\end{equation}
Here $\dot S_i$ parameterizes entropy production associated with irreversible processes or effective interactions while also providing a phenomenological handle on deviations that cannot be absorbed into $S_{\rm cos}$ or $T_{\rm eff}$. In either reading the recovery of the GR Friedmann equations from horizon thermodynamics is revealed as conditional. With the black hole inputs one obtains the GR relations because one has effectively inserted the GR content at the outset. With the TSC inputs one obtains a calibrated deformation that reduces to GR in the appropriate limit.
\\
\\
In practice one expects the influence of $F(H)$ or of $\dot S_i$ to be constrained by CMB peak physics, BAO, supernovae, and lensing. At the same time small early-time deformations that scale faster than radiation can shift the sound horizon and thus the inferred $H_0$ can also shift, while late-time deformations can reduce the amplitude of structure growth and hence impact $S_8$. The empirical status of such deformations depends on a global fit for sure, but the main point is conceptual rather than phenomenological. The TSC diagnoses that the path from black hole thermodynamics to cosmological dynamics is not set in stone and that the correct route runs through cosmology-native thermodynamic laws, with GR recovered as a limit and well posed quantum gravity corrections can appear as controlled additional terms in the background dynamics.
\\
\\
It is instructive to now make explicit how TSC naturally points to controlled deformations of the background dynamics that appear as small, quantitative departures from the Einstein–Friedmann relations while reducing to GR in the BH and GH limit. Note that when  $(\chi,p,\alpha)=(1,2,8\pi^2 M_{\rm Pl}^2)$ and $\dot S_i=0$ one recovers $\dot H=-4\pi G(\rho+p)$ identically in \eqref{fr1}-\eqref{fr2}. For generic $(\chi,p,\alpha)$ and small $\dot S_i$, deviations are controlled by dimensionless numbers of order a few per cent and by powers of $H$ fixed by $p$. A convenient formulation of the corresponding first Friedmann relation is obtained by integrating the generalized $\dot H$ along with using the continuity equation $\dot\rho+3H(\rho+p)=0$. One can then write a caliberated deformation of $\Lambda$CDM as follows
\begin{equation}
H^2(z) \;=\; H_0^2\,E_{\Lambda{\rm CDM}}^2(z) \;+\; F\!\left(H(z);\alpha,p,\chi\right)
\end{equation}
where $E_{\Lambda{\rm CDM}}^2(z)=\Omega_{r0}(1+z)^4+\Omega_{m0}(1+z)^3+\Omega_{\Lambda 0}$ and $F$ vanishes in the BH–GH limit. For phenomenology it is useful to adopt a small amplitude expansion around $\Lambda$CDM, either as some sort of quasi fluid or as a series in H. A simple ansatz consistent with the TSC scaling is
\begin{equation}
F(H) = \varepsilon_b \left(\frac{H}{H_b}\right)^{m} H_0^2
\end{equation}
with $H_b$ a pivot and $m$ determined by the entropy exponent $p$ and the near–dS factor $\chi$. In the equilibrium reading with $\dot S_i=0$, one finds $m=p$ for deformations that mimic a horizon–anchored contribution. The constant $\varepsilon_b$ is constrained at the percent level. If $m>2$ the additional density dilutes faster than radiation in the far past and can act as a localized early–time injection that raises $H(z)$ at $z\sim 10^3$ without spoiling BBN. If $m\simeq 2$ it behaves like a running vacuum term and If $m<2$, it is a late–time deformation that modifies acceleration and growth.
\\
\\
The sound horizon at the baryon drag epoch is given as 
\begin{equation}
r_s \;=\; \int_{z_d}^{\infty} \frac{c_s(z)}{H(z)}\,dz
\end{equation}
with $c_s$ being the photon–baryon sound speed. A positive $F$ that is non–negligible for $z\geq 10^3$ increases $H(z)$ in the integrand, thereby decreasing $r_s$. Since BAO will calibrate the product $H_0 r_s$ through the low–redshift distance ladder, a smaller $r_s$ implies a larger inferred $H_0$ at fixed BAO distances. A viable TSC–driven resolution of the $H_0$ tension therefore corresponds to a small, localized early time $F$ with $m\gtrsim 2$ and $|\varepsilon_b|\sim \mathrm{few}\times 10^{-2}$, raising $H(z)$ by a few per cent near recombination but dying away quickly at higher $z$ to preserve BBN and the damping tail of the CMB. The same deformation must be small at $z\lesssim 10^2$ so as not to overly accelerate the universe or distort the late–time distance ladder.
\\
\\
Another avenue of interest here would be the linear growth of matter perturbations, which is governed by
\begin{equation}
\ddot{\delta}_m + 2H\dot{\delta}_m - 4\pi G_{\rm eff}(a,k)\,\rho_m\,\delta_m \;=\; 0
\end{equation}
where $G_{\rm eff}$ reduces to $G$ in GR but can be renormalized by horizon–thermodynamic departures if one insists on an equilibrium mapping. When considering a background only viewpoint the friction term $2H\dot\delta_m$ is enhanced whenever $H$ is raised by a positive $F$ at late times, suppressing growth and lowering $S_8=\sigma_8 \left(\Omega_{m0}/0.3\right)^{1/2}$. This gives a very natural route to easing the $S_8$ tension by a small, smooth late–time deformation with $m\lesssim 2$ and $|\varepsilon_b|\sim 10^{-2}$, which increases $H(z)$ in the range $0\lesssim z\lesssim 1$ by order per cent and reduces the accumulated growth factor $D(z)$ accordingly. In the non–equilibrium reading, a positive entropy production rate can come as a effective bulk viscous pressure $\Pi=-3\zeta H$ at the fluid level and that being
\begin{equation}
\dot\rho_m + 3H\,\rho_m \;=\; 0, \qquad \dot\rho_X + 3H\left(1+w_X\right)\rho_X \;=\; 9\zeta H^2
\end{equation}
where the right hand side terms source an emergent component that can be tuned to produce the same small enhancements in $H(z)$ while also modifying the Euler equation for perturbations through an effective sound speed or viscosity, further suppressing the growth rate $f=d\ln D/d\ln a$.
\\
\\
It is also very useful to display explicitly the mapping between the TSC exponents and the quasi–fluid description. Suppose $F(H)=\varepsilon_b (H/H_b)^m H_0^2$ with $|\varepsilon_b|\ll 1$ then the effective energy density associated to $F$ is $\rho_F=(3/8\pi G)\,F$ and its adiabatic equation of state is obtained by differentiating $F$ with respect to $\ln a$
\begin{equation}
w_F(a) \;=\; -1 \;-\; \frac{1}{3}\,\frac{d\ln F}{d\ln a} \;=\; -1 \;+\; \frac{m}{3}\,\frac{d\ln H}{d\ln a}
\end{equation}
During radiation domination we know that $d\ln H/d\ln a \simeq -2$ and $w_F \simeq -1 + \tfrac{2m}{3}$ while during matter domination $d\ln H/d\ln a \simeq -\tfrac{3}{2}$ and $w_F \simeq -1 + \tfrac{m}{2}$. This tells us that a deformation with $m\approx 4$ behaves as a stiff, rapidly decaying injection at very early times and is safe for BBN, while $m\approx 3$ mimics an "early dark energy" \cite{ede1Poulin:2023lkg,ede2Cicoli:2023qri,ede3Kamionkowski:2022pkx,ede4McDonough:2023qcu,ede5Niedermann:2023ssr} like scaling that is non–negligible near recombination but quickly shuts off. Converse to this, $m\approx 1$ gives a late–time component whose EOS parameter is close to $w\simeq -\tfrac{1}{2}$ across matter domination and primarily impacts low redshift growth and distances. These scalings arise naturally from the entropy exponent $p$ through $m=p$ in the equilibrium mapping, so that measuring $p$ with thermodynamic probes directly fixes the phenomenology of $F$.
\\
\\
The entropic non–equilibrium term admits an equally transparent parametrization too as we can write $\dot S_i = \sigma(H)\,H^{-p-1}$ with a small, positive function $\sigma(H)$ to obtain
\begin{equation}
\dot H \;=\; -\,\frac{8\pi^2}{\chi\,p\,\alpha}\,(\rho+p)\,H^{\,p-2} \;+\; \frac{\sigma(H)}{p\,\alpha}
\end{equation}
so that $\sigma$ acts like a mild, explicitly time–dependent acceleration source that can be arranged to be non–zero only in a narrow redshift band. In this representation the impact on $H(z)$, $r_s$, and $D(z)$ is localized and can be dialed to be of order per cent. Because $\sigma$ is tied to an entropy production rate, it is also automatically positive and thus consistent with a generalized second law provided $\chi>0$ and $\alpha>0$, which are independently testable thermodynamic requirements. These are possibilities which we shall discuss in detail in upcoming papers.
\\
\\
All such deformations can be tightly constrained by the CMB acoustic structure \cite{Planck:2018vyg}, big bang nucleosynthesis \cite{bbn1Sarkar:1995dd,bbn2Fields:2019pfx}, BAO \cite{baoFerreira:2025lrd}, supernovae \cite{sup1Scolnic:2021amr,sup2DES:2024jxu}, lensing \cite{len1Bartelmann:2010fz,len2Ng:2024dcj}, RSD \cite{rsd1Hamilton:1997zq,rsd2Briffa:2023ozo,rsd3Paillas:2021oli}, and peculiar–velocity measurements \cite{pvs1Boubel:2023mfe,pvs2Turner:2024blz,pvs3Tsagas:2025pxi}. The point emphasized by the TSC is that once the cosmology–native horizon law is used, the GR Friedmann equations reappears as a limit of a broader class in which quantum gravity information is compressed into the empirical functions $\chi(H)$, $p$, $\alpha$, and into the production rate $\dot S_i$. The resulting corrections need not be large indeed but the data demand that they be subtle but also sufficient in a sense that percent level, redshift localized enhancements of $H(z)$ are exactly what is required to reduce the sound horizon enough to reconcile BAO with a larger $H_0$, and gentle late–time increases in $H(z)$ or small effective viscosities are precisely what is needed to lower the growth amplitude and ease the $S_8$ discrepancy. Within the TSC framework these are not ad hoc fluids or arbitrary functions but controlled, thermodynamically motivated terms whose scaling follows from the measured entropy exponent and effective temperature of cosmological horizons.
\\
\\
We would also like to note that the conflict exposed by the TSC is not a clash between GR and observation but between a black hole inspired thermodynamic dictionary and the structural realities of cosmology. General relativity as a local dynamical theory remains firmly intact, with its wide range of precision tests in the laboratory the solar system, binary pulsars and gravitational waves unchallenged. What shifts is the thermodynamic framework that was presumed to extend from black holes to cosmological horizons. If future measurements establish that $\chi\ne 1$ or $p\ne 2$, then the natural interpretation is that the universe follows a thermodynamic law distinct from that of black holes that we are accustomed to. The horizon Clausius rule then leads either to non-equilibrium production terms or to equilibrium deformations that manifest as additional H dependent terms in the Friedmann equations. Both readings are windows into quantum gravity because the empirical functions $S_{\rm cos}(H)$ and $T_{\rm eff}(H)$ summarize in the infrared, the accumulated effect of UV microphysics on long-wavelength gravitational thermodynamics which is quite a unique correspondence. The Euclidean partition function and specific heats, the cosmological island notions, and the horizon first derivations of FRW dynamics can then be recast using the same empirical inputs, restoring internal consistency once the correct boundary functional is known.
\\
\\
\section{Conclusions}
This particular work deals with quite a few interesting facets on the interface of quantum gravity, thermodynamics, and cosmology. By questioning the foundational assumption that the Gibbons–Hawking temperature and the Bekenstein–Hawking entropy directly apply to cosmological horizons, we have done a thorough re-examination of how the thermodynamic structure of the universe itself may be an empirical question rather than an inherited axiom. The Thermodynamic Split Conjecture provides a conceptual and mathematical framework for describing this departure and we would just like to summarize the main points which we have uncovered in our analysis here as follows:
\begin{itemize}
    \item Our work shows that the GH temperature and the area–law entropy while mathematically elegant, rely on structural features unique to black holes encapsulated in the BKE criterion which cosmological horizons generally lack. As such, the corresponding thermodynamic quantities for the universe must be reformulated in a cosmology–native form, $T_{\rm eff}(H)$ and $S_{\rm cos}(H)$, to ensure physical consistency.
    \item By deriving general, exponent–free forms for key cosmological relations, ranging from the Langevin dynamics of inflationary perturbations to the generalized Friedmann equation, we have demonstrated that all major early–universe processes (eternal inflation, Hawking–Moss transitions, PBH formation etc.) would depend strongly on the empirically consistent thermodynamic identities for cosmology.
    \item The work also clarifies that small deviations from the standard horizon thermodynamics could produce measurable corrections to cosmological evolution, potentially offering new ways to alleviate observational tensions such as the $H_0$ and $S_8$ discrepancies, without invoking exotic new physics or radical departures from general relativity. So maybe, one does not need to completely revamp general relativity to alleviate these issues but just has to understand thermodynamics better. 
    \item A promising direction for future works could be to connect the TSC framework to specific models of quantum gravity, for instance to string inspired or loop quantum cosmology scenarios, and identify how microscopic gravitational degrees of freedom manifest as corrections to $S_{\rm cos}(H)$ or $T_{\rm eff}(H)$ and can be probed through cosmological observables.
    \item Finally, the results presented here motivate the creation of a broader observational program aimed at testing all thermodynamic laws in a cosmological context, spanning energy conservation, horizon entropy growth, and temperature–entropy relations. This could end up laying the foundation for a dedicated mission too, which would be conceived to empirically verify the thermal constitution of the universe itself.
\end{itemize}
\section*{Acknowledgments}
The work of the author is supported with the Vanderbilt Discovery Doctoral Fellowship. The author would like to thank Avi Loeb, Robert Scherrer, Jan Novak ,Sunny Vagnozzi and Meet Vyas for helpful discussions. In particular the author would like to thank Avi as his suggestion to probe the implications of TSC on the early universe led to this work. The author would also like to thank Harvard's Center for Astrophysics for their hospitality where a part of this work was completed. 
\bibliography{references}

\begin{thebibliography}{100}

\bibitem{de1SupernovaSearchTeam:1998fmf}
Adam~G. Riess et~al.
\newblock {Observational evidence from supernovae for an accelerating universe and a cosmological constant}.
\newblock {\em Astron. J.}, 116:1009--1038, 1998.

\bibitem{de2Li:2012dt}
Miao Li, Xiao-Dong Li, Shuang Wang, and Yi~Wang.
\newblock {Dark Energy: A Brief Review}.
\newblock {\em Front. Phys. (Beijing)}, 8:828--846, 2013.

\bibitem{de3Li:2011sd}
Miao Li, Xiao-Dong Li, Shuang Wang, and Yi~Wang.
\newblock {Dark Energy}.
\newblock {\em Commun. Theor. Phys.}, 56:525--604, 2011.

\bibitem{de4Mortonson:2013zfa}
Michael~J. Mortonson, David~H. Weinberg, and Martin White.
\newblock {Dark Energy: A Short Review}.
\newblock 12 2013.

\bibitem{de5Frusciante:2019xia}
Noemi Frusciante and Louis Perenon.
\newblock {Effective field theory of dark energy: A review}.
\newblock {\em Phys. Rept.}, 857:1--63, 2020.

\bibitem{de6Huterer:2017buf}
Dragan Huterer and Daniel~L Shafer.
\newblock {Dark energy two decades after: Observables, probes, consistency tests}.
\newblock {\em Rept. Prog. Phys.}, 81(1):016901, 2018.

\bibitem{de7Vagnozzi:2021quy}
Sunny Vagnozzi, Luca Visinelli, Philippe Brax, Anne-Christine Davis, and Jeremy Sakstein.
\newblock {Direct detection of dark energy: The XENON1T excess and future prospects}.
\newblock {\em Phys. Rev. D}, 104(6):063023, 2021.

\bibitem{de8Adil:2023ara}
Shahnawaz~A. Adil, Upala Mukhopadhyay, Anjan~A. Sen, and Sunny Vagnozzi.
\newblock {Dark energy in light of the early JWST observations: case for a negative cosmological constant?}
\newblock {\em JCAP}, 10:072, 2023.

\bibitem{de9Feleppa:2025clx}
Fabiano Feleppa, Gaetano Lambiase, and Sunny Vagnozzi.
\newblock {Imprints of screened dark energy on non-local quantum correlations}.
\newblock 8 2025.

\bibitem{de10DiValentino:2020evt}
Eleonora Di~Valentino, Stefano Gariazzo, Olga Mena, and Sunny Vagnozzi.
\newblock {Soundness of Dark Energy properties}.
\newblock {\em JCAP}, 07(07):045, 2020.

\bibitem{de11Nojiri:2010wj}
Shin'ichi Nojiri and Sergei~D. Odintsov.
\newblock {Unified cosmic history in modified gravity: from F(R) theory to Lorentz non-invariant models}.
\newblock {\em Phys. Rept.}, 505:59--144, 2011.

\bibitem{de12Nojiri:2006ri}
Shin'ichi Nojiri and Sergei~D. Odintsov.
\newblock {Introduction to modified gravity and gravitational alternative for dark energy}.
\newblock {\em eConf}, C0602061:06, 2006.

\bibitem{de13Trivedi:2023zlf}
Oem Trivedi.
\newblock {Recent Advances in Cosmological Singularities}.
\newblock {\em Symmetry}, 16(3):298, 2024.

\bibitem{de14Trivedi:2022svr}
Oem Trivedi and Maxim Khlopov.
\newblock {Singularity formation in asymptotically safe cosmology with inhomogeneous equation of state}.
\newblock {\em JCAP}, 11:007, 2022.

\bibitem{de15Trivedi:2024inb}
Oem Trivedi, Ayush Bidlan, and Paulo Moniz.
\newblock {Fractional holographic dark energy}.
\newblock {\em Phys. Lett. B}, 858:139074, 2024.

\bibitem{de16Trivedi:2024dju}
Oem Trivedi and Robert~J. Scherrer.
\newblock {New perspectives on future rip scenarios with holographic dark energy}.
\newblock {\em Phys. Rev. D}, 110(2):023521, 2024.

\bibitem{dm11rubin1970rotation}
Vera~C Rubin and W~Kent Ford~Jr.
\newblock Rotation of the andromeda nebula from a spectroscopic survey of emission regions.
\newblock {\em Astrophysical Journal, vol. 159, p. 379}, 159:379, 1970.

\bibitem{dm1Cirelli:2024ssz}
Marco Cirelli, Alessandro Strumia, and Jure Zupan.
\newblock {Dark Matter}.
\newblock 6 2024.

\bibitem{dm2Arbey:2021gdg}
A.~Arbey and F.~Mahmoudi.
\newblock {Dark matter and the early Universe: a review}.
\newblock {\em Prog. Part. Nucl. Phys.}, 119:103865, 2021.

\bibitem{dm3Balazs:2024uyj}
Csaba Balazs, Torsten Bringmann, Felix Kahlhoefer, and Martin White.
\newblock {A Primer on Dark Matter}.
\newblock 11 2024.

\bibitem{dm4Eberhardt:2025caq}
Andrew Eberhardt and Elisa G.~M. Ferreira.
\newblock {Ultralight fuzzy dark matter review}.
\newblock 7 2025.

\bibitem{dm5Bozorgnia:2024pwk}
Nassim Bozorgnia, Joseph Bramante, James~M. Cline, David Curtin, David McKeen, David~E. Morrissey, Adam Ritz, Simon Viel, Aaron~C. Vincent, and Yue Zhang.
\newblock {Dark Matter Candidates and Searches}.
\newblock 9 2024.

\bibitem{dm6Misiaszek:2023sxe}
Marcin Misiaszek and Nicola Rossi.
\newblock {Direct Detection of Dark Matter: A Critical Review}.
\newblock {\em Symmetry}, 16(2):201, 2024.

\bibitem{dm7OHare:2024nmr}
Ciaran A.~J. O'Hare.
\newblock {Cosmology of axion dark matter}.
\newblock {\em PoS}, COSMICWISPers:040, 2024.

\bibitem{dm8Adhikari:2022sbh}
Susmita Adhikari et~al.
\newblock {Astrophysical Tests of Dark Matter Self-Interactions}.
\newblock 7 2022.

\bibitem{dm9Miller:2025yyx}
Andrew~L. Miller.
\newblock {Gravitational wave probes of particle dark matter: a review}.
\newblock 3 2025.

\bibitem{dm10Trivedi:2025vry}
Oem Trivedi and Abraham Loeb.
\newblock {Could planck star remnants be dark matter?}
\newblock {\em Phys. Dark Univ.}, 49:102003, 2025.

\bibitem{ht1DiValentino:2021izs}
Eleonora Di~Valentino, Olga Mena, Supriya Pan, Luca Visinelli, Weiqiang Yang, Alessandro Melchiorri, David~F. Mota, Adam~G. Riess, and Joseph Silk.
\newblock {In the realm of the Hubble tension{\textemdash}a review of solutions}.
\newblock {\em Class. Quant. Grav.}, 38(15):153001, 2021.

\bibitem{ht2Clifton:2024mdy}
Timothy Clifton and Neil Hyatt.
\newblock {A radical solution to the Hubble tension problem}.
\newblock {\em JCAP}, 08:052, 2024.

\bibitem{s81kazantzidis2018evolution}
Lavrentios Kazantzidis and Leandros Perivolaropoulos.
\newblock Evolution of the f $\sigma$ 8 tension with the planck 15/$\lambda$ cdm determination and implications for modified gravity theories.
\newblock {\em Physical Review D}, 97(10):103503, 2018.

\bibitem{s82amon2022non}
Alexandra Amon and George Efstathiou.
\newblock A non-linear solution to the s 8 tension?
\newblock {\em Monthly Notices of the Royal Astronomical Society}, 516(4):5355--5366, 2022.

\bibitem{s83poulin2023sigma}
Vivian Poulin, Jos{\'e}~Luis Bernal, Ely~D Kovetz, and Marc Kamionkowski.
\newblock Sigma-8 tension is a drag.
\newblock {\em Physical Review D}, 107(12):123538, 2023.

\bibitem{s84Ferreira:2025lrd}
Elisa G.~M. Ferreira, Evan McDonough, Lennart Balkenhol, Renata Kallosh, Lloyd Knox, and Andrei Linde.
\newblock {The BAO-CMB Tension and Implications for Inflation}.
\newblock 7 2025.

\bibitem{baoFerreira:2025lrd}
Elisa G.~M. Ferreira, Evan McDonough, Lennart Balkenhol, Renata Kallosh, Lloyd Knox, and Andrei Linde.
\newblock {The BAO-CMB Tension and Implications for Inflation}.
\newblock 7 2025.

\bibitem{t1jha2019next}
Saurabh~W Jha, Federica Bianco, W~Niel Brandt, Gaspar Galaz, Eric Gawiser, John Gizis, Ren{\'e}e Hlo{\v{z}}ek, Sugata Kaviraj, Jeffrey~A Newman, Aprajita Verma, et~al.
\newblock Next generation lsst science.
\newblock {\em arXiv preprint arXiv:1907.08945}, 2019.

\bibitem{t2Shanks:2015lda}
T.~Shanks et~al.
\newblock {The VLT Survey Telescope ATLAS}.
\newblock {\em Mon. Not. Roy. Astron. Soc.}, 451(4):4238--4252, 2015.

\bibitem{t3chandler2025nsf}
Colin~Orion Chandler, Pedro~H Bernardinelli, Mario Juri{\'c}, Devanshi Singh, Henry~H Hsieh, Ian Sullivan, R~Lynne Jones, Jacob~A Kurlander, Dmitrii Vavilov, Siegfried Eggl, et~al.
\newblock Nsf-doe vera c. rubin observatory observations of interstellar comet 3i/atlas (c/2025 n1).
\newblock {\em arXiv preprint arXiv:2507.13409}, 2025.

\bibitem{t4Euclid:2024yrr}
Y.~Mellier et~al.
\newblock {Euclid. I. Overview of the Euclid mission}.
\newblock {\em Astron. Astrophys.}, 697:A1, 2025.

\bibitem{t5WST:2024rai}
Vincenzo Mainieri et~al.
\newblock {The Wide-field Spectroscopic Telescope (WST) Science White Paper}.
\newblock 3 2024.

\bibitem{t6CosmoVerseNetwork:2025alb}
Eleonora Di~Valentino et~al.
\newblock {The CosmoVerse White Paper: Addressing observational tensions in cosmology with systematics and fundamental physics}.
\newblock {\em Phys. Dark Univ.}, 49:101965, 2025.

\bibitem{t7COMPACT:2022gbl}
Yashar Akrami et~al.
\newblock {Promise of Future Searches for Cosmic Topology}.
\newblock {\em Phys. Rev. Lett.}, 132(17):171501, 2024.

\bibitem{tscpaper}
Oem Trivedi.
\newblock {Thermodynamic Split Conjecture and an Observational Test for Cosmological Entropy}.
\newblock 9 2025.

\bibitem{sb1hawking1972black}
Stephen~W Hawking.
\newblock Black holes in general relativity.
\newblock {\em Communications in Mathematical Physics}, 25:152--166, 1972.

\bibitem{sb3bekenstein1974generalized}
Jacob~D Bekenstein.
\newblock Generalized second law of thermodynamics in black-hole physics.
\newblock {\em Physical Review D}, 9(12):3292, 1974.

\bibitem{sb4Cotler:2016fpe}
Jordan~S. Cotler, Guy Gur-Ari, Masanori Hanada, Joseph Polchinski, Phil Saad, Stephen~H. Shenker, Douglas Stanford, Alexandre Streicher, and Masaki Tezuka.
\newblock {Black Holes and Random Matrices}.
\newblock {\em JHEP}, 05:118, 2017.
\newblock [Erratum: JHEP 09, 002 (2018)].

\bibitem{sb5krolak1978singularities}
A~Kr{\'o}lak.
\newblock {\em Singularities and black holes in general space-times}.
\newblock PhD thesis, MSc Thesis, 1978.

\bibitem{sb8joshi2002cosmic}
Pankaj~S Joshi.
\newblock Cosmic censorship: A current perspective.
\newblock {\em Modern Physics Letters A}, 17(15n17):1067--1079, 2002.

\bibitem{sb9senovilla20151965}
Jos{\'e}~MM Senovilla and David Garfinkle.
\newblock The 1965 penrose singularity theorem.
\newblock {\em Classical and Quantum Gravity}, 32(12):124008, 2015.

\bibitem{sb12hawking1970singularities}
Stephen~William Hawking and Roger Penrose.
\newblock The singularities of gravitational collapse and cosmology.
\newblock {\em Proceedings of the Royal Society of London. A. Mathematical and Physical Sciences}, 314(1519):529--548, 1970.

\bibitem{sb16janis1968reality}
Allen~I Janis, Ezra~T Newman, and Jeffrey Winicour.
\newblock Reality of the schwarzschild singularity.
\newblock {\em Physical Review Letters}, 20(16):878, 1968.

\bibitem{sv1Strominger:1996sh}
Andrew Strominger and Cumrun Vafa.
\newblock {Microscopic origin of the Bekenstein-Hawking entropy}.
\newblock {\em Phys. Lett. B}, 379:99--104, 1996.

\bibitem{sv2Nishioka:2009un}
Tatsuma Nishioka, Shinsei Ryu, and Tadashi Takayanagi.
\newblock {Holographic Entanglement Entropy: An Overview}.
\newblock {\em J. Phys. A}, 42:504008, 2009.

\bibitem{sv3Strominger:1997eq}
Andrew Strominger.
\newblock {Black hole entropy from near horizon microstates}.
\newblock {\em JHEP}, 02:009, 1998.

\bibitem{qe1Sen:2008vm}
Ashoke Sen.
\newblock {Quantum Entropy Function from AdS(2)/CFT(1) Correspondence}.
\newblock {\em Int. J. Mod. Phys. A}, 24:4225--4244, 2009.

\bibitem{sv4Bena:2005va}
Iosif Bena and Nicholas~P. Warner.
\newblock {Bubbling supertubes and foaming black holes}.
\newblock {\em Phys. Rev. D}, 74:066001, 2006.

\bibitem{sv5Carlip:1998wz}
Steven Carlip.
\newblock {Black hole entropy from conformal field theory in any dimension}.
\newblock {\em Phys. Rev. Lett.}, 82:2828--2831, 1999.

\bibitem{sv6DeHaro:2019gno}
Sebastian De~Haro, Jeroen van Dongen, Manus Visser, and Jeremy Butterfield.
\newblock {Conceptual analysis of black hole entropy in string theory}.
\newblock {\em Stud. Hist. Phil. Sci. B}, 69:82--111, 2020.

\bibitem{sv7Emparan:2006it}
Roberto Emparan and Gary~T. Horowitz.
\newblock {Microstates of a Neutral Black Hole in M Theory}.
\newblock {\em Phys. Rev. Lett.}, 97:141601, 2006.

\bibitem{car1Cardy:1986ie}
John~L. Cardy.
\newblock {Operator Content of Two-Dimensional Conformally Invariant Theories}.
\newblock {\em Nucl. Phys. B}, 270:186--204, 1986.

\bibitem{car2Cardy:1989ir}
John~L. Cardy.
\newblock {Boundary Conditions, Fusion Rules and the Verlinde Formula}.
\newblock {\em Nucl. Phys. B}, 324:581--596, 1989.

\bibitem{adscftMaldacena:1997re}
Juan~Martin Maldacena.
\newblock {The Large $N$ limit of superconformal field theories and supergravity}.
\newblock {\em Adv. Theor. Math. Phys.}, 2:231--252, 1998.

\bibitem{qe2Sen:2009vz}
Ashoke Sen.
\newblock {Arithmetic of Quantum Entropy Function}.
\newblock {\em JHEP}, 08:068, 2009.

\bibitem{qe3Sen:2012kpz}
Ashoke Sen.
\newblock {Logarithmic Corrections to N=2 Black Hole Entropy: An Infrared Window into the Microstates}.
\newblock {\em Gen. Rel. Grav.}, 44(5):1207--1266, 2012.

\bibitem{qe4Banerjee:2010qc}
Shamik Banerjee, Rajesh~Kumar Gupta, and Ashoke Sen.
\newblock {Logarithmic Corrections to Extremal Black Hole Entropy from Quantum Entropy Function}.
\newblock {\em JHEP}, 03:147, 2011.

\bibitem{qe5Banerjee:2011jp}
Shamik Banerjee, Rajesh~Kumar Gupta, Ipsita Mandal, and Ashoke Sen.
\newblock {Logarithmic Corrections to N=4 and N=8 Black Hole Entropy: A One Loop Test of Quantum Gravity}.
\newblock {\em JHEP}, 11:143, 2011.

\bibitem{qe6Sen:2012cj}
Ashoke Sen.
\newblock {Logarithmic Corrections to Rotating Extremal Black Hole Entropy in Four and Five Dimensions}.
\newblock {\em Gen. Rel. Grav.}, 44:1947--1991, 2012.

\bibitem{qe7Dabholkar:2014ema}
Atish Dabholkar, Joao Gomes, and Sameer Murthy.
\newblock {Nonperturbative black hole entropy and Kloosterman sums}.
\newblock {\em JHEP}, 03:074, 2015.

\bibitem{qe8Castro:2009jf}
Alejandra Castro and Finn Larsen.
\newblock {Near Extremal Kerr Entropy from AdS(2) Quantum Gravity}.
\newblock {\em JHEP}, 12:037, 2009.

\bibitem{nonst1sohl2015deep}
Jascha Sohl-Dickstein, Eric Weiss, Niru Maheswaranathan, and Surya Ganguli.
\newblock Deep unsupervised learning using nonequilibrium thermodynamics.
\newblock In {\em International conference on machine learning}, pages 2256--2265. pmlr, 2015.

\bibitem{nonst2gallavotti2019nonequilibrium}
Giovanni Gallavotti.
\newblock Nonequilibrium thermodynamics.
\newblock {\em arXiv preprint arXiv:1901.08821}, 2019.

\bibitem{nonst3Motta:2025xli}
Mario Motta, Antonio Mezzacapo, and Giacomo Guarnieri.
\newblock {Non-equilibrium thermodynamics of precision through a quantum-centric computation}.
\newblock 3 2025.

\bibitem{nonst4zwanzig2001nonequilibrium}
Robert Zwanzig.
\newblock {\em Nonequilibrium statistical mechanics}.
\newblock Oxford university press, 2001.

\bibitem{gh1Gibbons:1977mu}
G.~W. Gibbons and S.~W. Hawking.
\newblock {Cosmological Event Horizons, Thermodynamics, and Particle Creation}.
\newblock {\em Phys. Rev. D}, 15:2738--2751, 1977.

\bibitem{gh2Gibbons:1976ue}
G.~W. Gibbons and S.~W. Hawking.
\newblock {Action Integrals and Partition Functions in Quantum Gravity}.
\newblock {\em Phys. Rev. D}, 15:2752--2756, 1977.

\bibitem{gh3Gibbons:1978ac}
G.~W. Gibbons, S.~W. Hawking, and M.~J. Perry.
\newblock {Path Integrals and the Indefiniteness of the Gravitational Action}.
\newblock {\em Nucl. Phys. B}, 138:141--150, 1978.

\bibitem{gh4Hawking:1995fd}
S.~W. Hawking and Gary~T. Horowitz.
\newblock {The Gravitational Hamiltonian, action, entropy and surface terms}.
\newblock {\em Class. Quant. Grav.}, 13:1487--1498, 1996.

\bibitem{ska1Weltman:2018zrl}
A.~Weltman et~al.
\newblock {Fundamental physics with the Square Kilometre Array}.
\newblock {\em Publ. Astron. Soc. Austral.}, 37:e002, 2020.

\bibitem{ska2Maartens:2015mra}
Roy Maartens, Filipe~B. Abdalla, Matt Jarvis, and Mario~G. Santos.
\newblock {Overview of Cosmology with the SKA}.
\newblock {\em PoS}, AASKA14:016, 2015.

\bibitem{ska3Santos:2015gra}
Mario~G. Santos et~al.
\newblock {Cosmology from a SKA HI intensity mapping survey}.
\newblock {\em PoS}, AASKA14:019, 2015.

\bibitem{hera1DeBoer:2016tnn}
David~R. DeBoer et~al.
\newblock {Hydrogen Epoch of Reionization Array (HERA)}.
\newblock {\em Publ. Astron. Soc. Pac.}, 129(974):045001, 2017.

\bibitem{hera2HERA:2021noe}
Zara Abdurashidova et~al.
\newblock {HERA Phase I Limits on the Cosmic 21 cm Signal: Constraints on Astrophysics and Cosmology during the Epoch of Reionization}.
\newblock {\em Astrophys. J.}, 924(2):51, 2022.

\bibitem{et2gibbons1983very}
Gary~W Gibbons, Stephen~W Hawking, and Stephen~TC Siklos.
\newblock The very early universe: proceedings of the nuffield workshop, cambridge, 21 june to 9 july, 1982.
\newblock {\em Very Early Universe}, 1983.

\bibitem{et3vilenkin1983birth}
Alexander Vilenkin.
\newblock Birth of inflationary universes.
\newblock {\em Physical Review D}, 27(12):2848, 1983.

\bibitem{et4guth1985quantum}
Alan~H Guth and So-Young Pi.
\newblock Quantum mechanics of the scalar field in the new inflationary universe.
\newblock {\em Physical Review D}, 32(8):1899, 1985.

\bibitem{et5vanchurin2000predictability}
Vitaly Vanchurin, Alexander Vilenkin, and Serge Winitzki.
\newblock Predictability crisis in inflationary cosmology and its resolution.
\newblock {\em Physical Review D}, 61(8):083507, 2000.

\bibitem{et6linde1986eternal}
Andrei~D Linde.
\newblock Eternal chaotic inflation.
\newblock {\em Modern Physics Letters A}, 1(02):81--85, 1986.

\bibitem{et7linde1986eternally}
Andrei~D Linde.
\newblock Eternally existing self-reproducing chaotic inflanationary universe.
\newblock {\em Physics Letters B}, 175(4):395--400, 1986.

\bibitem{et8goncharov1987global}
AS~Goncharov, Andrei~D Linde, and Viatcheslav~F Mukhanov.
\newblock The global structure of the inflationary universe.
\newblock {\em International Journal of Modern Physics A}, 2(03):561--591, 1987.

\bibitem{et9trivedi2022rejuvenating}
Oem Trivedi.
\newblock Rejuvenating the hope of a swampland consistent inflated multiverse with tachyonic inflation in the high-energy rs-ii braneworld.
\newblock {\em Modern Physics Letters A}, 37(24):2250162, 2022.

\bibitem{et10guth2000inflation}
Alan~H Guth.
\newblock Inflation and eternal inflation.
\newblock {\em Physics Reports}, 333:555--574, 2000.

\bibitem{et1guth2007eternal}
Alan~H Guth.
\newblock Eternal inflation and its implications.
\newblock {\em Journal of Physics A: Mathematical and Theoretical}, 40(25):6811, 2007.

\bibitem{hm1hawking1982supercooled}
Stephen~William Hawking and Ian~L Moss.
\newblock Supercooled phase transitions in the very early universe.
\newblock {\em Physics Letters B}, 110(1):35--38, 1982.

\bibitem{hm2coleman1980gravitational}
Sidney Coleman and Frank De~Luccia.
\newblock Gravitational effects on and of vacuum decay.
\newblock {\em Physical Review D}, 21(12):3305, 1980.

\bibitem{hm3gregory2020black}
Ruth Gregory, Ian~G Moss, and Naritaka Oshita.
\newblock Black holes, oscillating instantons and the hawking-moss transition.
\newblock {\em Journal of High Energy Physics}, 2020(7):1--22, 2020.

\bibitem{hm4coleman1977fate}
Sidney Coleman.
\newblock Fate of the false vacuum: Semiclassical theory.
\newblock {\em Physical Review D}, 15(10):2929, 1977.

\bibitem{hm5callan1977fate}
Curtis~G Callan~Jr and Sidney Coleman.
\newblock Fate of the false vacuum. ii. first quantum corrections.
\newblock {\em Physical Review D}, 16(6):1762, 1977.

\bibitem{br1dvali2017quantum}
Gia Dvali, Cesar Gomez, and Sebastian Zell.
\newblock Quantum break-time of de sitter.
\newblock {\em Journal of Cosmology and Astroparticle Physics}, 2017(06):028, 2017.

\bibitem{tcc1Bedroya:2019snp}
Alek Bedroya and Cumrun Vafa.
\newblock {Trans-Planckian Censorship and the Swampland}.
\newblock {\em JHEP}, 09:123, 2020.

\bibitem{tcc2Bedroya:2019tba}
Alek Bedroya, Robert Brandenberger, Marilena Loverde, and Cumrun Vafa.
\newblock {Trans-Planckian Censorship and Inflationary Cosmology}.
\newblock {\em Phys. Rev. D}, 101(10):103502, 2020.

\bibitem{tcc3Andriot:2020lea}
David Andriot, Niccol{\`o} Cribiori, and David Erkinger.
\newblock {The web of swampland conjectures and the TCC bound}.
\newblock {\em JHEP}, 07:162, 2020.

\bibitem{tcc4Brahma:2020zpk}
Suddhasattwa Brahma, Omar Alaryani, and Robert Brandenberger.
\newblock {Entanglement entropy of cosmological perturbations}.
\newblock {\em Phys. Rev. D}, 102(4):043529, 2020.

\bibitem{tcc5Blumenhagen:2019vgj}
Ralph Blumenhagen, Max Brinkmann, and Andriana Makridou.
\newblock {Quantum Log-Corrections to Swampland Conjectures}.
\newblock {\em JHEP}, 02:064, 2020.

\bibitem{tcc6Mizuno:2019bxy}
Shuntaro Mizuno, Shinji Mukohyama, Shi Pi, and Yun-Long Zhang.
\newblock {Universal Upper Bound on the Inflationary Energy Scale from the Trans-Planckian Censorship Conjecture}.
\newblock {\em Phys. Rev. D}, 102(2):021301, 2020.

\bibitem{tcc7Brahma:2019unn}
Suddhasattwa Brahma.
\newblock {Trans-Planckian censorship, inflation and excited initial states for perturbations}.
\newblock {\em Phys. Rev. D}, 101(2):023526, 2020.

\bibitem{tcc8Berera:2019zdd}
Arjun Berera and Jaime~R. Calder{\'o}n.
\newblock {Trans-Planckian censorship and other swampland bothers addressed in warm inflation}.
\newblock {\em Phys. Rev. D}, 100(12):123530, 2019.

\bibitem{tcc9Cai:2019hge}
Yong Cai and Yun-Song Piao.
\newblock {Pre-inflation and trans-Planckian censorship}.
\newblock {\em Sci. China Phys. Mech. Astron.}, 63(11):110411, 2020.

\bibitem{pbh10Belotsky:2018wph}
Konstantin~M. Belotsky, Vyacheslav~I. Dokuchaev, Yury~N. Eroshenko, Ekaterina~A. Esipova, Maxim~Yu. Khlopov, Leonid~A. Khromykh, Alexander~A. Kirillov, Valeriy~V. Nikulin, Sergey~G. Rubin, and Igor~V. Svadkovsky.
\newblock {Clusters of primordial black holes}.
\newblock {\em Eur. Phys. J. C}, 79(3):246, 2019.

\bibitem{pbh2Carr:1974nx}
Bernard~J. Carr and S.~W. Hawking.
\newblock {Black holes in the early Universe}.
\newblock {\em Mon. Not. Roy. Astron. Soc.}, 168:399--415, 1974.

\bibitem{pbh3Carr:1975qj}
Bernard~J. Carr.
\newblock {The Primordial black hole mass spectrum}.
\newblock {\em Astrophys. J.}, 201:1--19, 1975.

\bibitem{pbh4Carr:2020gox}
Bernard Carr, Kazunori Kohri, Yuuiti Sendouda, and Jun'ichi Yokoyama.
\newblock {Constraints on primordial black holes}.
\newblock {\em Rept. Prog. Phys.}, 84(11):116902, 2021.

\bibitem{pbh5Page:1976df}
Don~N. Page.
\newblock {Particle Emission Rates from a Black Hole: Massless Particles from an Uncharged, Nonrotating Hole}.
\newblock {\em Phys. Rev. D}, 13:198--206, 1976.

\bibitem{pbh6Carr:2016drx}
Bernard Carr, Florian Kuhnel, and Marit Sandstad.
\newblock {Primordial Black Holes as Dark Matter}.
\newblock {\em Phys. Rev. D}, 94(8):083504, 2016.

\bibitem{pbh7Khlopov:2008qy}
Maxim~Yu. Khlopov.
\newblock {Primordial Black Holes}.
\newblock {\em Res. Astron. Astrophys.}, 10:495--528, 2010.

\bibitem{pbh8Khlopov:1985fch}
M.~Yu. Khlopov, B.~A. Malomed, Ia.~B. Zeldovich, and Ya.~B. Zeldovich.
\newblock {Gravitational instability of scalar fields and formation of primordial black holes}.
\newblock {\em Mon. Not. Roy. Astron. Soc.}, 215(4):575--589, 1985.

\bibitem{pbh9Khlopov:1980mg}
M.~Yu. Khlopov and A.~G. Polnarev.
\newblock {PRIMORDIAL BLACK HOLES AS A COSMOLOGICAL TEST OF GRAND UNIFICATION}.
\newblock {\em Phys. Lett. B}, 97:383--387, 1980.

\bibitem{ha1Hayward:1997jp}
Sean~A. Hayward.
\newblock {Unified first law of black hole dynamics and relativistic thermodynamics}.
\newblock {\em Class. Quant. Grav.}, 15:3147--3162, 1998.

\bibitem{ha2Hayward:2008jq}
S.~A. Hayward, R.~Di~Criscienzo, L.~Vanzo, M.~Nadalini, and S.~Zerbini.
\newblock {Local Hawking temperature for dynamical black holes}.
\newblock {\em Class. Quant. Grav.}, 26:062001, 2009.

\bibitem{Cai:2005ra}
Rong-Gen Cai and Sang~Pyo Kim.
\newblock {First law of thermodynamics and Friedmann equations of Friedmann-Robertson-Walker universe}.
\newblock {\em JHEP}, 02:050, 2005.

\bibitem{z1Govindarajan:2002ry}
T.~R. Govindarajan, R.~K. Kaul, and V.~Suneeta.
\newblock {Quantum gravity on dS(3)}.
\newblock {\em Class. Quant. Grav.}, 19:4195--4205, 2002.

\bibitem{is1Hartman:2020khs}
Thomas Hartman, Yikun Jiang, and Edgar Shaghoulian.
\newblock {Islands in cosmology}.
\newblock {\em JHEP}, 11:111, 2020.

\bibitem{is2Azarnia:2021uch}
Sanam Azarnia, Reza Fareghbal, Ali Naseh, and Hamed Zolfi.
\newblock {Islands in flat-space cosmology}.
\newblock {\em Phys. Rev. D}, 104(12):126017, 2021.

\bibitem{is3Espindola:2022fqb}
Ricardo Esp{\'\i}ndola, Bahman Najian, and Dora Nikolakopoulou.
\newblock {Islands in FRW Cosmologies}.
\newblock 3 2022.

\bibitem{is4Bousso:2022gth}
Raphael Bousso and Elizabeth Wildenhain.
\newblock {Islands in closed and open universes}.
\newblock {\em Phys. Rev. D}, 105(8):086012, 2022.

\bibitem{is5Ben-Dayan:2022nmb}
Ido Ben-Dayan, Merav Hadad, and Elizabeth Wildenhain.
\newblock {Islands in the fluid: islands are common in cosmology}.
\newblock {\em JHEP}, 03:077, 2023.

\bibitem{therm1Khodam-Mohammadi:2023ndu}
A.~Khodam-Mohammadi and M.~Monshizadeh.
\newblock {Exploring modifications to FLRW cosmology with general entropy and thermodynamics: A new approach}.
\newblock {\em Phys. Lett. B}, 843:138066, 2023.

\bibitem{therm2Nojiri:2022aof}
Shin'ichi Nojiri, Sergei~D. Odintsov, and Valerio Faraoni.
\newblock {From nonextensive statistics and black hole entropy to the holographic dark universe}.
\newblock {\em Phys. Rev. D}, 105(4):044042, 2022.

\bibitem{therm3Akbar:2006kj}
M.~Akbar and Rong-Gen Cai.
\newblock {Thermodynamic Behavior of Friedmann Equations at Apparent Horizon of FRW Universe}.
\newblock {\em Phys. Rev. D}, 75:084003, 2007.

\bibitem{therm4Bargueno:2021nuc}
Pedro Bargue{\~n}o, Ernesto Contreras, and {\'A}ngel Rinc{\'o}n.
\newblock {Thermodynamics of scale-dependent Friedmann equations}.
\newblock {\em Eur. Phys. J. C}, 81(5):477, 2021.

\bibitem{therm5Sheykhi:2010zz}
Ahmad Sheykhi.
\newblock {Thermodynamics of apparent horizon and modified Friedmann equations}.
\newblock {\em Eur. Phys. J. C}, 69:265--269, 2010.

\bibitem{therm6Helou:2015yqa}
Alexis Helou.
\newblock {Dynamics of the Cosmological Apparent Horizon: Surface Gravity {\&} Temperature}.
\newblock 2 2015.

\bibitem{therm7Nojiri:2019skr}
Shin'ichi Nojiri, Sergei~D. Odintsov, and Emmanuel~N. Saridakis.
\newblock {Modified cosmology from extended entropy with varying exponent}.
\newblock {\em Eur. Phys. J. C}, 79(3):242, 2019.

\bibitem{therm8Nojiri:2022nmu}
Shin'ichi Nojiri, Sergei~D. Odintsov, and Tanmoy Paul.
\newblock {Modified cosmology from the thermodynamics of apparent horizon}.
\newblock {\em Phys. Lett. B}, 835:137553, 2022.

\bibitem{therm9Nojiri:2022dkr}
Shin'ichi Nojiri, Sergei~D. Odintsov, and Tanmoy Paul.
\newblock {Early and late universe holographic cosmology from a new generalized entropy}.
\newblock {\em Phys. Lett. B}, 831:137189, 2022.

\bibitem{therm10Odintsov:2022qnn}
Sergei~D. Odintsov and Tanmoy Paul.
\newblock {A non-singular generalized entropy and its implications on bounce cosmology}.
\newblock {\em Phys. Dark Univ.}, 39:101159, 2023.

\bibitem{therm11Nojiri:2023bom}
Shin'ichi Nojiri, Sergei~D. Odintsov, and Tanmoy Paul.
\newblock {Microscopic interpretation of generalized entropy}.
\newblock {\em Phys. Lett. B}, 847:138321, 2023.

\bibitem{therm12Nojiri:2024zdu}
Shin'ichi Nojiri, Sergei~D. Odintsov, and Tanmoy Paul.
\newblock {Different Aspects of Entropic Cosmology}.
\newblock {\em Universe}, 10(9):352, 2024.

\bibitem{therm13Nojiri:2023ikl}
Shin'ichi Nojiri and Sergei~D. Odintsov.
\newblock {Micro-canonical and canonical description for generalised entropy}.
\newblock {\em Phys. Lett. B}, 845:138130, 2023.

\bibitem{therm14deHaro:2023lbq}
Jaume de~Haro, Shin'ichi Nojiri, S.~D. Odintsov, V.~K. Oikonomou, and Supriya Pan.
\newblock {Finite-time cosmological singularities and the possible fate of the Universe}.
\newblock {\em Phys. Rept.}, 1034:1--114, 2023.

\bibitem{therm15Brevik:2024ozg}
I.~Brevik, Maxim Khlopov, S.~D. Odintsov, Alexander~V. Timoshkin, and Oem Trivedi.
\newblock {Rips and regular future scenario with holographic dark energy: a comprehensive look}.
\newblock {\em Eur. Phys. J. C}, 84(12):1269, 2024.

\bibitem{ede1Poulin:2023lkg}
Vivian Poulin, Tristan~L. Smith, and Tanvi Karwal.
\newblock {The Ups and Downs of Early Dark Energy solutions to the Hubble tension: A review of models, hints and constraints circa 2023}.
\newblock {\em Phys. Dark Univ.}, 42:101348, 2023.

\bibitem{ede2Cicoli:2023qri}
Michele Cicoli, Matteo Licheri, Ratul Mahanta, Evan McDonough, Francisco~G. Pedro, and Marco Scalisi.
\newblock {Early Dark Energy in Type IIB String Theory}.
\newblock {\em JHEP}, 06:052, 2023.

\bibitem{ede3Kamionkowski:2022pkx}
Marc Kamionkowski and Adam~G. Riess.
\newblock {The Hubble Tension and Early Dark Energy}.
\newblock {\em Ann. Rev. Nucl. Part. Sci.}, 73:153--180, 2023.

\bibitem{ede4McDonough:2023qcu}
Evan McDonough, J.~Colin Hill, Mikhail~M. Ivanov, Adrien La~Posta, and Michael~W. Toomey.
\newblock {Observational constraints on early dark energy}.
\newblock {\em Int. J. Mod. Phys. D}, 33(11):2430003, 2024.

\bibitem{ede5Niedermann:2023ssr}
Florian Niedermann and Martin~S. Sloth.
\newblock {New Early Dark Energy as a solution to the $H_0$ and $S_8$ tensions}.
\newblock 7 2023.

\bibitem{Planck:2018vyg}
N.~Aghanim et~al.
\newblock {Planck 2018 results. VI. Cosmological parameters}.
\newblock {\em Astron. Astrophys.}, 641:A6, 2020.
\newblock [Erratum: Astron.Astrophys. 652, C4 (2021)].

\bibitem{bbn1Sarkar:1995dd}
Subir Sarkar.
\newblock {Big bang nucleosynthesis and physics beyond the standard model}.
\newblock {\em Rept. Prog. Phys.}, 59:1493--1610, 1996.

\bibitem{bbn2Fields:2019pfx}
Brian~D. Fields, Keith~A. Olive, Tsung-Han Yeh, and Charles Young.
\newblock {Big-Bang Nucleosynthesis after Planck}.
\newblock {\em JCAP}, 03:010, 2020.
\newblock [Erratum: JCAP 11, E02 (2020)].

\bibitem{sup1Scolnic:2021amr}
Dan Scolnic et~al.
\newblock {The Pantheon+ Analysis: The Full Data Set and Light-curve Release}.
\newblock {\em Astrophys. J.}, 938(2):113, 2022.

\bibitem{sup2DES:2024jxu}
T.~M.~C. Abbott et~al.
\newblock {The Dark Energy Survey: Cosmology Results with {\ensuremath{\sim}}1500 New High-redshift Type Ia Supernovae Using the Full 5 yr Data Set}.
\newblock {\em Astrophys. J. Lett.}, 973(1):L14, 2024.

\bibitem{len1Bartelmann:2010fz}
Matthias Bartelmann.
\newblock {Gravitational Lensing}.
\newblock {\em Class. Quant. Grav.}, 27:233001, 2010.

\bibitem{len2Ng:2024dcj}
Angela L.~H. Ng.
\newblock {Cosmology using Strong Gravitational Lensing}.
\newblock Other thesis, 5 2024.

\bibitem{rsd1Hamilton:1997zq}
A.~J.~S. Hamilton.
\newblock {Linear redshift distortions: A Review}.
\newblock In {\em {Ringberg Workshop on Large Scale Structure}}, 8 1997.

\bibitem{rsd2Briffa:2023ozo}
Rebecca Briffa, Celia Escamilla-Rivera, Jackson Levi~Said, and Jurgen Mifsud.
\newblock {Growth of structures using redshift space distortion in f(T) cosmology}.
\newblock {\em Mon. Not. Roy. Astron. Soc.}, 528(2):2711--2727, 2024.

\bibitem{rsd3Paillas:2021oli}
Enrique Paillas, Yan-Chuan Cai, Nelson Padilla, and Ariel~G. S{\'a}nchez.
\newblock {Redshift-space distortions with split densities}.
\newblock {\em Mon. Not. Roy. Astron. Soc.}, 505(4):5731--5752, 2021.

\bibitem{pvs1Boubel:2023mfe}
Paula Boubel, Matthew Colless, Khaled Said, and Lister Staveley-Smith.
\newblock {Large-scale motions and growth rate from forward-modelling Tully{\textendash}Fisher peculiar velocities}.
\newblock {\em Mon. Not. Roy. Astron. Soc.}, 531(1):84--109, 2024.

\bibitem{pvs2Turner:2024blz}
Ryan~J. Turner.
\newblock {Cosmology with Peculiar Velocity Surveys}.
\newblock 11 2024.

\bibitem{pvs3Tsagas:2025pxi}
Christos~G. Tsagas, Leandros Perivolaropoulos, and Kerkyra Asvesta.
\newblock {Large-scale peculiar velocities in the universe}.
\newblock 10 2025.

\end{thebibliography}
\bibliographystyle{unsrt}

\end{document}